\documentclass[reprint,prl,footinbib,floatfix,twocolumn,longbibliography]{revtex4-2}
\usepackage{times}
\usepackage{amssymb}
\usepackage{color}
\usepackage{amsmath}
\usepackage{amsbsy}
\usepackage{amsthm}
\usepackage{graphicx}
\usepackage{bbm}
\usepackage{bm}
\usepackage{epsfig}
\usepackage{xfrac}
\usepackage{xcolor}
\usepackage{enumerate}
\usepackage{multirow}
\usepackage{physics}
\usepackage[T2A,T1]{fontenc}
\usepackage{ tipa }
\usepackage[english]{babel}
\usepackage{symbols}
\usepackage{makecell}
\usepackage{appendix}
\usepackage{dsfont}
\usepackage{float}
\usepackage{pdfpages}
\usepackage{bbold}
\usepackage{placeins}
\usepackage{relsize}
\usepackage{xcolor}
\usepackage{scalerel}
\usepackage{tikz}
\usetikzlibrary{svg.path}

\DeclareMathSymbol{\shortminus}{\mathbin}{AMSa}{"39}

\definecolor{orcidlogocol}{HTML}{A6CE39}
\tikzset{
  orcidlogo/.pic={
    \fill[orcidlogocol] svg{M256,128c0,70.7-57.3,128-128,128C57.3,256,0,198.7,0,128C0,57.3,57.3,0,128,0C198.7,0,256,57.3,256,128z};
    \fill[white] svg{M86.3,186.2H70.9V79.1h15.4v48.4V186.2z}
                 svg{M108.9,79.1h41.6c39.6,0,57,28.3,57,53.6c0,27.5-21.5,53.6-56.8,53.6h-41.8V79.1z M124.3,172.4h24.5c34.9,0,42.9-26.5,42.9-39.7c0-21.5-13.7-39.7-43.7-39.7h-23.7V172.4z}
                 svg{M88.7,56.8c0,5.5-4.5,10.1-10.1,10.1c-5.6,0-10.1-4.6-10.1-10.1c0-5.6,4.5-10.1,10.1-10.1C84.2,46.7,88.7,51.3,88.7,56.8z};
  }
}

\newcommand\orcid[1]{\href{https://orcid.org/#1}{$\,$\mbox{\scalerel*{
\begin{tikzpicture}[yscale=-1,transform shape]
\pic{orcidlogo};
\end{tikzpicture}
}{|}}}}

\definecolor{myurlcolor}{rgb}{0.0,0.39,0.0}
\definecolor{myrefcolor}{rgb}{0.0,0.39,0.0}

\usepackage[colorlinks]{hyperref}
\hypersetup{unicode=true,
    bookmarksopen=false,
    breaklinks=false,
    pdfborder={0 0 0},
	bookmarksnumbered=false,
	pdfstartview={FitH},
	citecolor={myurlcolor},
	linkcolor={myrefcolor},
	urlcolor={myurlcolor}}

\usepackage{soul}

\makeatletter
\AtBeginDocument{\let\LS@rot\@undefined}
\makeatother 

\makeatletter
\def\maketitle{
\@author@finish
\title@column\titleblock@produce
\suppressfloats[t]}
\makeatother

\makeatletter
\newcommand{\smtableofcontents}{%
  \section*{Contents}%
  \@starttoc{smtoc}%
}
\newcommand{\smsection}[1]{%
  \section{#1}%
  \addcontentsline{smtoc}{section}{\protect\numberline{\thesection}#1}%
}
\newcommand{\smsubsection}[1]{%
  \subsection{#1}%
  \addcontentsline{smtoc}{subsection}{\protect\numberline{\thesubsection}#1}%
}
\newcommand{\smsubsubsection}[1]{%
  \subsubsection{#1}%
  \addcontentsline{smtoc}{subsubsection}{\protect\numberline{\thesubsubsection}#1}%
}
\makeatother

\begin{document}

\title{To Cool, or Not to Cool? Displacement Sensing with Hot Quantum States}

\author{Piotr~T.~Grochowski\orcid{0000-0002-9654-4824}}
\email{piotr.grochowski@upol.cz}
\affiliation{Department of Optics, \href{https://ror.org/04qxnmv42}{Palacký University}, 17. listopadu 1192/12, 771 46 Olomouc, Czech Republic}

\begin{abstract}
Quantum-enhanced displacement sensing with bosonic systems is typically formulated assuming that the oscillator is cooled close to its ground state before nonclassical probe preparation.
We investigate whether such near-ground-state initialization is necessary, or whether sensitive probes can instead be generated directly from thermal states.
We analyze hot quantum probes produced by squeezing, number-raising, and Schrödinger-cat-state generation applied to thermal inputs.
We identify two distinct mechanisms by which thermal mixedness can remain compatible with enhanced displacement sensitivity.
First, projecting a mixed probe onto a definite parity sector removes the usual thermal suppression of the displacement quantum Fisher information, which can then increase with initial thermal occupation.
Second, coherent superpositions of opposite displacements can retain sensitivity through coherence between their displaced components, even when the underlying state is mixed.
We use these two mechanisms to classify hot-state protocols according to whether their sensitivity comes from parity selection, coherence between displaced components, or both.
Finally, we formulate an experimentally relevant optimization problem comparing initial cooling with direct hot-state preparation under realistic decoherence and show that complete cooling is not universally optimal.
Our results establish hot-state engineering as a route to quantum-enhanced bosonic displacement sensing without mandatory ground-state initialization.
\end{abstract}

\maketitle

\textit{Introduction}---Quantum sensing with bosonic systems has emerged as a powerful framework for precision measurements using engineered continuous-variable quantum states~\cite{Paris2009,Degen2017,Pirandola2018,Fadel2025}.
In this setting, sensitivity enhancements arise from nonclassical bosonic resources~\cite{Lachman2022,Walschaers2021,Rakhubovsky2024,Frigerio2025}, including squeezed states, Fock states~\cite{Chu2018,Wolf2019,Podhora2022,Hofheinz2008,Eickbusch2022,Deng2024,Rahman2025}, their superpositions~\cite{McCormick2019,Wang2019,Kovalenko2025}, and Schr\"odinger cat states~\cite{Vlastakis2013,Pan2025,Zheng2026a}, whose phase-space interference structure enables sensitivities beyond those achievable with coherent or thermal probes~\cite{Braun2018,Duivenvoorden2017,Fadel2025,Toscano2006,Grochowski2025}.
Rapid progress across optical~\cite{Rouviere2024}, superconducting~\cite{Eickbusch2022,Deng2024,Pan2025,Zheng2026a}, atomic~\cite{Facon2016,Burd2019,Wolf2019,Gilmore2021,Delakouras2023,Valahu2025,Bond2025}, and mechanical~\cite{Gavartin2012,Rakhubovsky2024,Satzinger2018,Mason2019,Wollack2022,vonLupke2022,Bild2023,Youssefi2023,Millen2020,Gonzalez-Ballestero2021} platforms has established such states as key resources for quantum sensing.
However, most existing displacement-sensing protocols rely on near-ground-state cooling~\cite{Degen2017,Aspelmeyer2014}, reflecting the common expectation that thermal occupation destroys the phase-space structures responsible for enhanced sensitivity~\cite{Zurek2001,Toscano2006}.
Still, it remains unclear to what extent displacement-sensitive structures can survive in bosonic systems initialized far in the thermal regime.

\begin{figure}[ht!]
    \includegraphics[width=\linewidth]{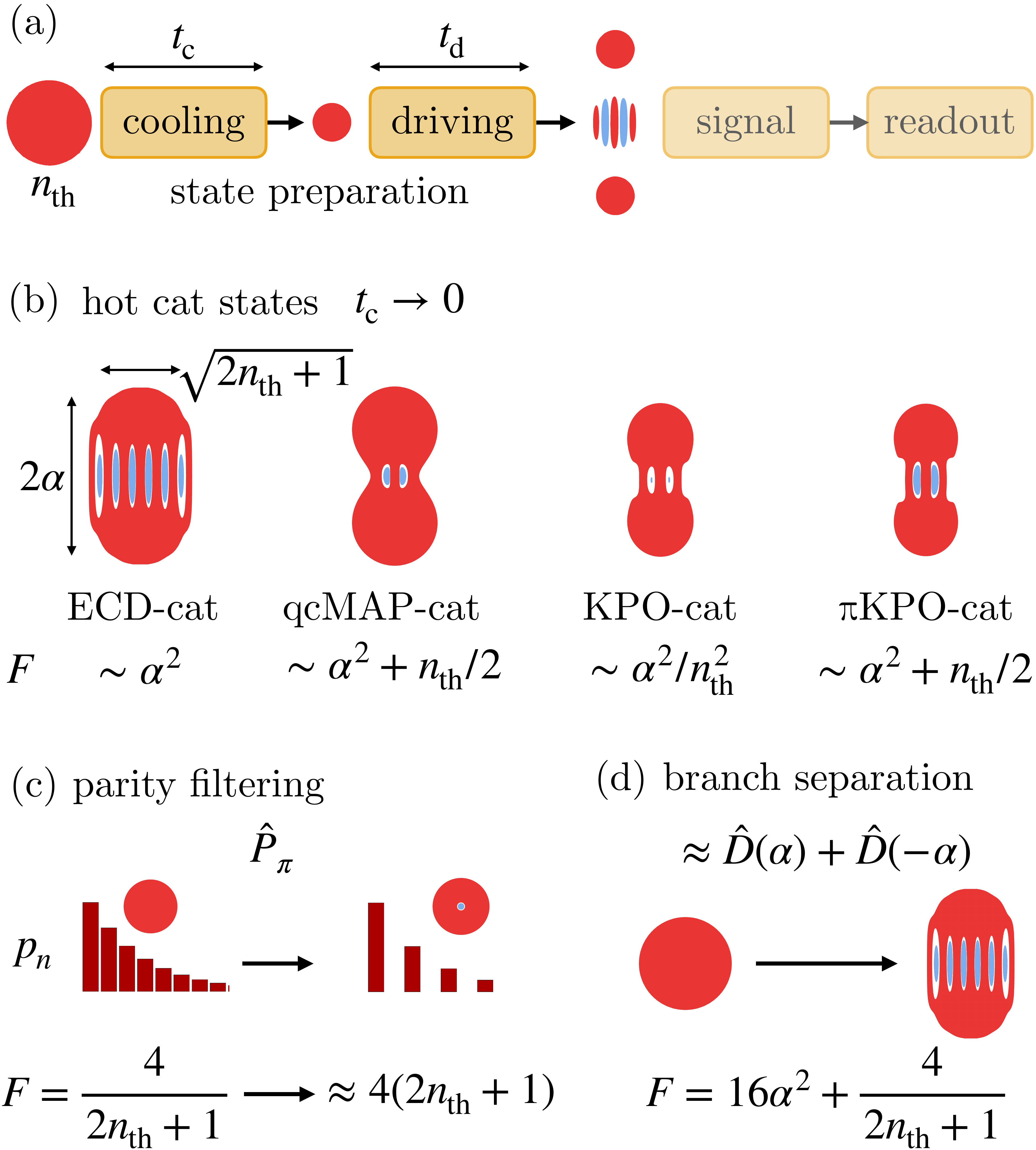}  
    \caption{Operational structure of hot-state bosonic sensing and representative resources.
(a) Cycle: cooling, nonclassical-state preparation, signal accumulation, and readout.
(b) Hot states are generated by applying a nonclassical-state preparation protocol directly to a thermal rather than ground state.
Echoed conditional displacement (ECD), qubit-cavity mapping (qcMAP), and Kerr-parametric-oscillator (KPO) protocols generate distinct hot-cat classes with different thermal-sensitivity scaling.
Here $\pi$KPO denotes KPO catification preceded by parity filtering.
These scalings depend on two mechanisms.
(c) Projection onto a definite parity sector leaves the thermal population in one sector, while the displacement generator couples it to the empty opposite sector, changing the asymptotic scaling from $\QFisher=4/(2\ThermalOccupation+1)$ to $\QFisher\approx 4(2\ThermalOccupation+1)$.
(d) Approximate superpositions of opposite displacements retain a contribution from phase-space coherence, even when the underlying state is mixed.
  \label{Fig1}}
\end{figure}

From an operational perspective, bosonic sensing involves a full cycle of cooling, nonclassical-state preparation, signal accumulation, and readout [cf. Fig.~\ref{Fig1}(a)].
Each stage consumes time and experimental resources under constraints set by decoherence, finite repetition rates, and limited control fidelity.
Since ground-state cooling may itself be demanding in many platforms~\cite{Leibfried2003,Clerk2010,Wineland2013,Aspelmeyer2014,Degen2017,Millen2020,Gonzalez-Ballestero2021}, it is natural to ask whether the optimal strategy is always to cool first, or whether useful displacement sensitivity can be engineered directly from a thermal input.

Here we show that this is indeed possible.
We first introduce the operational paradigm of hot-state preparation and then identify two qualitatively distinct mechanisms that preserve large displacement sensitivity within thermal ensembles: parity-sector engineering and coherent branch interference between displaced thermal components.
We use these mechanisms to classify several families of hot metrological states and analyze their noise fragility.
Finally, we consider an experimentally realistic competition between cooling and direct hot-state preparation and show that the latter can outperform full cooling-based sensing cycles.

\textit{Hot quantum states}---Standard nonclassical-state preparation begins from a bosonic mode cooled close to its ground state.
A hot-state protocol instead applies the same preparation operation directly to a thermal input, thereby avoiding near-ground-state initialization as part of the sensing cycle [see Fig.~\ref{Fig1}].
The recent realization of hot Schr\"odinger cat states~\cite{Yang2025} has shown that nonclassical phase-space structure can survive this procedure even at low initial purity~\cite{Zhu1996,Huyet2001,Jeong2006,Zheng2007,Jeong2007,Nicacio2010}.
Motivated by this, we consider hot squeezed states and hot Fock states generated through excitation addition~\cite{Zavatta2004,Parigi2007,Zavatta2007}, Jaynes--Cummings (JC) ladder climbing~\cite{Meekhof1996,Brune1996,Law1996,Hofheinz2008}, or Susskind--Glogower (SG) shifts~\cite{Susskind1964,Krastanov2015,Heeres2015,Heeres2017}.
We also consider hot cat states generated through echoed conditional displacement (ECD)~\cite{Eickbusch2022,Yang2025}, qubit-cavity mapping (qcMAP)~\cite{Leghtas2013a,Yang2025}, or Kerr-parametric-oscillator (KPO) catification~\cite{Puri2020,Frattini2024,Frattini2021}.
These operations generally increase the energy and may produce visible nonclassical structure, but neither property alone guarantees large mixed-state quantum Fisher information.
The central question is therefore which structural features prevent thermal spectral mixing from suppressing displacement sensitivity.
We identify two such mechanisms below.

\textit{Parity-sector engineering}---Consider a continuous-variable bosonic mode, where $\Creation$ adds a single excitation and $[\Annihilation,\Creation]=1$.
We focus on sensing a phase-space displacement generated by $\Quadrature=\Annihilation+\Creation$.
For a probe state $\DensityMatrix=\sum_n\Prob_n\ket{\EigenVal_n}\bra{\EigenVal_n}$, the displaced family reads $\DensityMatrix_\DisplaceSense=\text{exp}(-\ImagUnit\DisplaceSense\Quadrature)\DensityMatrix \, \text{exp}(\ImagUnit\DisplaceSense\Quadrature)$.
The sensitivity is bounded by the Cram\'er--Rao theorem~\cite{Holevo2011}, $\Delta\tilde{\DisplaceSense}\geq1/[\sqrt{\NoRuns\QFisher(\DensityMatrix_\DisplaceSense)}]$, where $\NoRuns$ is the number of repetitions and the quantum Fisher information (QFI)~\cite{Braunstein1994,Paris2009} is
\begin{align}
    \QFisher\rounds{\DensityMatrix_\DisplaceSense}
    =2\sum_{m,n}\frac{\rounds{\Prob_m-\Prob_n}^2}{\Prob_m+\Prob_n}
    \straights{\bra{\EigenVal_m}\Quadrature\ket{\EigenVal_n}}^2.
    \label{eq:qfi_general}
\end{align}
Typically, thermal occupation suppresses displacement sensitivity despite increasing the excitation number.
For instance, a thermal state yields $\QFisher=4\Purity=4/(2\ThermalOccupation+1)$, where $\ThermalOccupation$ is the thermal occupation and $\Purity$ the purity.

The simplest realization of parity-sector engineering is parity filtering.
Let $\Projector_{\pm}=(1\pm\ParityOp)/2$, where $\ParityOp=\exp(\ImagUnit\pi\Creation\Annihilation)$.
The filtered state $\DensityMatrix_{\pm}=\Projector_{\pm}\DensityMatrix\Projector_{\pm}/\text{Tr}(\Projector_{\pm}\DensityMatrix\Projector_{\pm})$ has support in a definite-parity sector.
Since displacement flips parity, $\ParityOp\Quadrature\ParityOp=-\Quadrature$, the generator couples the occupied sector exclusively to the initially empty opposite-parity sector, $\Projector_{\pm}\Quadrature\Projector_{\pm}=0$.
Then Eq.~\eqref{eq:qfi_general} simplifies to
\begin{align}
    \QFisher
    =4\Tr\DensityMatrix_{\pm}\Quadrature^2
    =4\rounds{2\angles{\Creation\Annihilation}_{\pm}+1
    +2\Re\angles{\Annihilation^2}_{\pm}}.
    \label{eq:qfi_parity_only}
\end{align}
This simplification is closely related to general mixed-state QFI enhancement when the signal generator couples the occupied support to orthogonal sectors~\cite{Petz2011,Paris2009}.
For example, for an even-parity-filtered thermal state, $\QFisher=4[4\ThermalOccupation^2/(2\ThermalOccupation+1)+1]\approx4/\Purity$ for $\ThermalOccupation\gg1$, reversing the scaling with purity.
Parity filtering therefore converts thermal occupation from a source of spectral suppression into a source of metrological sensitivity.
Notably, related enhancement has been found for measurement-projected thermal spin ensembles~\cite{Tatsuta2019,Tatsuta2026}.
The same parity-sector structure is closely related to bosonic cat-code encodings, where photon loss appears as a parity-flip error~\cite{Mirrahimi2014}. This parity-sector structure also implies vulnerability to parity-breaking noise, since loss or diffusion populates the initially empty sector, as analyzed below.

\textit{Coherent branch engineering}---Parity filtering is sufficient to remove thermal spectral suppression, but it is not necessary.
A fundamentally different mechanism is realized by hot ECD-cats, which retain large displacement sensitivity without definite-parity preparation.
Starting from a thermal bosonic state in a dispersively coupled spin--boson system, the ECD protocol coherently applies opposite bosonic displacements and erases the which-branch information stored in the ancilla by the final qubit projection.
In the large-separation regime, $\CohDisp^2\gg2\ThermalOccupation+1$, this generates the approximately normalized mixture
$\HotCatState_\ECDLetter\simeq\sum_n\Prob_n\ket{n,\CohDisp}\bra{n,\CohDisp}$, where
$\ket{n,\CohDisp}=[\Displacement{\CohDisp}\ket n+\Displacement{-\CohDisp}\ket n]/\sqrt2$.

To expose the resulting structure, we introduce an approximate tensor product between an effective branch pseudospin and local intrabranch fluctuations:
$\ket{\pm,n}=\Displacement{\pm\CohDisp}\ket n\simeq\ket{\pm}_{\mathrm b}\ket n_{\mathrm{int}}$.
Within this displaced subspace,
$\Annihilation=\CohDisp\PauliZ^{\mathrm b}+\Annihilation_{\mathrm{int}}+\mathcal O(e^{-2\CohDisp^2})$,
where $\PauliZ^{\mathrm b}\ket{\pm}_{\mathrm b}=\pm\ket{\pm}_{\mathrm b}$.
The ECD-cat then separates into coherent and thermal parts,
\begin{align}
    \HotCatState_\ECDLetter
    \simeq
    \ket{\times}_{\mathrm b}\bra{\times}
    \otimes\sum_n\Prob_n\ket n_{\mathrm{int}}\bra n,
\end{align}
where $\ket{\times}_{\mathrm b}=(\ket+_{\mathrm b}+\ket-_{\mathrm b})/\sqrt2$ is the balanced coherent branch superposition.
Because both the state and the generator separate at leading order, the QFI becomes
\begin{align}
    \QFisher
    &= 4 {\rm{Var}}(2 \CohDisp\PauliZ^{{\mathrm{b}}})_{\times}
    + \QFisher[\ThermalState,\Quadrature_{\mathrm{int}}]
    = 16\CohDisp^2 + \frac{4}{2\ThermalOccupation+1},
    \label{eq:fisher_ecd}
\end{align}
with $\Quadrature_{\mathrm{int}}=\Annihilation_{\mathrm{int}}+\Creation_{\mathrm{int}}$.
Eq.~\eqref{eq:fisher_ecd} separates the metrological enhancement into coherent branch and local thermal contributions.
The branch term remains available because the same relative cat phase is imprinted on every thermally populated Fock component.
An incoherent mixture of the two branches would eliminate it.
Thermal mixedness is instead confined to intrabranch fluctuations, while the dominant displacement-sensitive contribution is carried by coherence between the displaced branches.
This structure is closely related to the branch-interference picture underlying macroscopic cat-state sensitivity and phase-space sub-Planck structures~\cite{Zurek2001,Toscano2006,Frowis2018}, but with coherence encoded between displaced thermal branches rather than locally pure states.
It differs fundamentally from parity enhancement: parity-filtered protocols rely on Hilbert-space exclusion between opposite parity ladders, whereas ECD-cats store the resource in an approximately pure branch degree of freedom.

\textit{Classification of hot-state protocols}---The two mechanisms above organize the metrological behavior of the hot-state families introduced earlier.
For each protocol, whenever physically meaningful, we compare the standard hot-state preparation applied directly to the full thermal ensemble with a parity-filtered variant in which one parity sector is selected before preparation.
Parity projection has long appeared in quantum sensing and bosonic quantum information, primarily at the readout or logical-encoding level~\cite{Gerry2000,Anisimov2010,Vlastakis2013,Sun2014a,Ofek2016}.
Here, it instead structures the prepared thermal probe itself.
The resulting asymptotic QFI scalings are summarized in Table~\ref{tab:qfi_summary}, with detailed derivations given in SM~\cite{suppMat6}.

Ordinary hot squeezing and number-state raising generally fail to retain large QFI because thermal occupation produces spectral overlap between neighboring parity sectors despite increasing the excitation number.
Parity filtering removes this suppression for squeezed, excitation-added, and SG-shifted states.
For JC ladder climbing, however, the calibrated dynamics itself repopulates the opposite parity sector, so the parity enhancement is not generally preserved for deep ladders.
Among hot cats, qcMAP and parity-selected KPO states derive their enhancement from a definite-parity structure, whereas ECD realizes coherent branch engineering without explicit parity filtering.
Ordinary KPO cats contain coherent branch contributions, but thermal averaging over the initial even and odd sectors suppresses them.
Thus, the metrological behavior is determined not by energy or visible nonclassicality alone, but by how the preparation protocol structures the thermal ensemble.

\textit{Noise susceptibility of hot states}---The physical structure carrying the metrological enhancement also determines how different noise channels degrade it.
During the sensing or storage stage, we consider bosonic loss, motional heating, and bosonic number dephasing,
\begin{align}
    \partial_\DimTime\DensityMatrix
    &=
    \DimBosonLossRate\Dissipator{\Annihilation}\DensityMatrix
    +
    \DimHeatingRate
    \rounds{
        \Dissipator{\Annihilation}
        +
        \Dissipator{\Creation}
    }
    \DensityMatrix
    +
    \DimBosonDephasingRate
    \Dissipator{\Creation\Annihilation}\DensityMatrix .
\end{align}
Here, $\Dissipator{\hat O}\DensityMatrix=\hat O\DensityMatrix\hat O^\dagger-\{\hat O^\dagger\hat O,\DensityMatrix\}/2$ denotes the Lindblad dissipator~\cite{Breuer2007}.
Loss and heating transfer population between opposite parity sectors, whereas number dephasing preserves parity.
As a measure of the initial fragility, we use the instantaneous susceptibility $\DotQFisher=\left.\partial_\DimTime\QFisher[\DensityMatrix(\DimTime),\Quadrature]\right|_{\DimTime=0}$.
General information-theoretic bounds based on the quantum Stam inequality and its attenuation-channel counterpart~\cite{DePalma2017} imply $\DotQFisher\leq-\DimHeatingRate\QFisher^2/2$ for heating and $\DotQFisher\leq-\DimBosonLossRate\QFisher(\QFisher-4)/4$ for loss, as shown in SM~\cite{suppMat6}.
Thus, at large $\QFisher$, both channels enforce an initial degradation whose magnitude grows at least quadratically with the displacement sensitivity, up to channel-dependent prefactors.
The analytical results and full Lindblad simulations in SM~\cite{suppMat6} show how this general sensitivity--fragility relation is realized for the different hot-state resources.

\begin{table}[t]
\caption{
Asymptotic displacement QFI for the considered hot-state protocols.
Here, $\CohDisp$ is cat-state displacement, $\Squeeze$ is the squeezing parameter, $\ResPar=\ThermalOccupation/(1+\ThermalOccupation)$, $\ThermalOccupation$ is the initial thermal occupation of the state, $k$ is the number of added excitations in hot-Fock schemes, and PF denotes parity filtering. The JC-Fock PF scaling refers to shallow ladders before calibrated JC dynamics repopulates the opposite parity sector.
 }
\begin{ruledtabular}
\begin{tabular}{lcc}
State & no PF & with PF \\
\hline

sq. thermal &
$4e^{2\Squeeze}/{(2 \ThermalOccupation+1)}$ &
$\sim 8e^{2\Squeeze}\ThermalOccupation$
\\

add-Fock &
$2/(1 + \ThermalOccupation)$ &
$\sim 8(k+1)\ThermalOccupation$
\\

JC-Fock &
$4(2k+1)/(\ThermalOccupation+1)$ &
$\sim \ThermalOccupation$
\\

SG-Fock &
$\sim (4k+2)/\ThermalOccupation$ &
$\sim 8 \ThermalOccupation +8k$
\\

ECD-cat &
$\displaystyle
16\CohDisp^2+\frac{4}{2 \ThermalOccupation +1}
$
&
N/A
\\

qcMAP-cat &
N/A &
$\displaystyle
16\CohDisp^2+4(2 \ThermalOccupation +1)
$
\\

KPO-cat &
$\displaystyle
\frac{16\CohDisp^2}{(2 \ThermalOccupation+1)^2}
+
4\frac{1-\ResPar^2}{1+\ResPar^2}
$
&
$\displaystyle
16\CohDisp^2+4(2 \ThermalOccupation +1)
$

\end{tabular}
\end{ruledtabular}
\label{tab:qfi_summary}
\end{table}

For parity-engineered states, loss and heating cause leakage into the initially empty parity sector.
For branch-dominated cats, they additionally reveal which-branch information and suppress coherence between the displaced branches~\cite{Pan2023}.
Both mechanisms can coexist in parity-selected cat states.
Number dephasing behaves differently because it preserves parity.
Under number dephasing, initially Fock-diagonal parity-filtered resources have $\DotQFisher=0$, whereas states whose QFI relies on phase-sensitive squeezing or coherent cat branches remain susceptible~\cite{suppMat6}.

Full Lindblad simulations at approximately matched initial QFI further show that the robustness ordering is channel dependent (reported in full in SM~\cite{suppMat6}).
Among the considered cat states, ECD-cats are the most robust under loss and heating, parity-filtered KPO-cats show intermediate robustness, and ordinary KPO-cats are particularly fragile under loss.
Fock-diagonal parity-filtered resources are instead the most robust against number dephasing~\cite{suppMat6}.
Consequently, there is no universal robustness hierarchy.
At fixed initial QFI, the decay depends on which structure carries the sensitivity, how the noise acts on that structure, and the energetic cost required to generate it.
Depending on the specific experimental platform and its intrinsic decoherence channels, different hot states may prove beneficial.

\textit{To cool, or not to cool?}---The existence of highly sensitive hot states raises an operational question: does extensive pre-cooling remain advantageous once the full sensing cycle is taken into account?
A realistic protocol consists of cooling, state preparation, signal accumulation, and readout, with total cycle duration
$\DimCycleTime=\DimCoolTime+\DimCatTime+\DimTimeSense+\DimTimeMeas$.
Since the number of repetitions during a fixed experimental time scales as $\NoRuns\simeq\FullDimTime/\DimCycleTime$, the relevant operational figure of merit is the sensitivity generated per unit total time,
$\DimProtocolRate=\QFisher/\DimCycleTime$, consistent with treating both time and energy as resources in noisy bosonic metrology~\cite{Gorecki2025}.
This figure of merit applies to independent repetitions of a fixed local-estimation protocol.
For adaptive or sequential strategies, Bayesian risk or conditional information gain per total time may be more appropriate~\cite{Fiderer2021,OConnor2024}.

We analyze $\DimProtocolRate$ for ECD-cat preparation.
Unlike the preceding sensing-stage analysis, the preparation dynamics acts on a joint spin--boson system.
The branch coherence is therefore reduced by ancilla spin dephasing, described by the additional dissipator $(\DimSpinDephasingRate/4)\Dissipator{\PauliZ}$, and by the dominant platform-specific bosonic channel.
As an example, we focus below on trapped ions, where this channel is the same motional heating considered above, while the corresponding circuit-QED expression with photon loss is given in SM~\cite{suppMat6}.

The key observation is that the dominant branch-interference term in Eq.~\eqref{eq:fisher_ecd} depends primarily on the cat separation and only weakly on the initial thermal occupation.
Cooling therefore increases the experimental overhead while affecting only the subleading local contribution, whereas preparation-stage decoherence suppresses the dominant branch coherence.
We use dimensionless times $\Time=\JCCoup\DimTime$, where $\JCCoup$ is the spin--boson coupling strength.
The dimensionless cooling rate $\CoolingRate=\DimCoolingRate/\JCCoup$ describes relaxation toward a bath with negligible residual occupation, so that $\PostCoolingOccupation(\CoolTime)=\ThermalOccupation\exp(-\CoolingRate\CoolTime)$.
The dimensionless rates $\SpinDephasingRate=\DimSpinDephasingRate/\JCCoup$ and $\HeatingRate=\DimHeatingRate/\JCCoup$ describe ancilla spin dephasing and motional heating during the subsequent ECD preparation.
During ECD generation, the branch displacement grows as $\CohDisp(\CatTime)=\CatTime$.
For trapped ions, the large-separation QFI is approximated by
\begin{align}
    \QFisher
    \simeq
    16\CatTime^2\CatVisibility^2(\CatTime)
    +
    \frac{4}{2\PostCoolingOccupation(\CoolTime)+1},
    \label{eq:dec_simp}
\end{align}
with $\CatVisibility^2=\exp[-\SpinDephasingRate\CatTime-(8/3)\HeatingRate\CatTime^3]$, where the two factors describe ancilla spin dephasing and motional heating, respectively (see SM for derivation~\cite{suppMat6}).
For analytical transparency, we use the simplified objective $\QFisher/(\CoolTime+\CatTime)$, justified when $\TimeSense+\TimeMeas$ is negligible or fixed and sufficiently small not to change the optimum qualitatively.
Whenever $16\CatTime^2\CatVisibility^2\gg4/(2\PostCoolingOccupation+1)$, additional cooling produces only a minor QFI increase while substantially lengthening the protocol.
In this branch-dominated regime, the optimization yields $\CoolTime^\star\simeq0$, together with a finite optimal preparation time.
When spin dephasing dominates, $\Time_{\mathrm{d}}^\star=1/\SpinDephasingRate$.
When motional heating dominates, $\Time_{\mathrm{d}}^\star=(8\HeatingRate)^{-1/3}$.

\begin{figure}[t!]
\includegraphics[width=\linewidth]{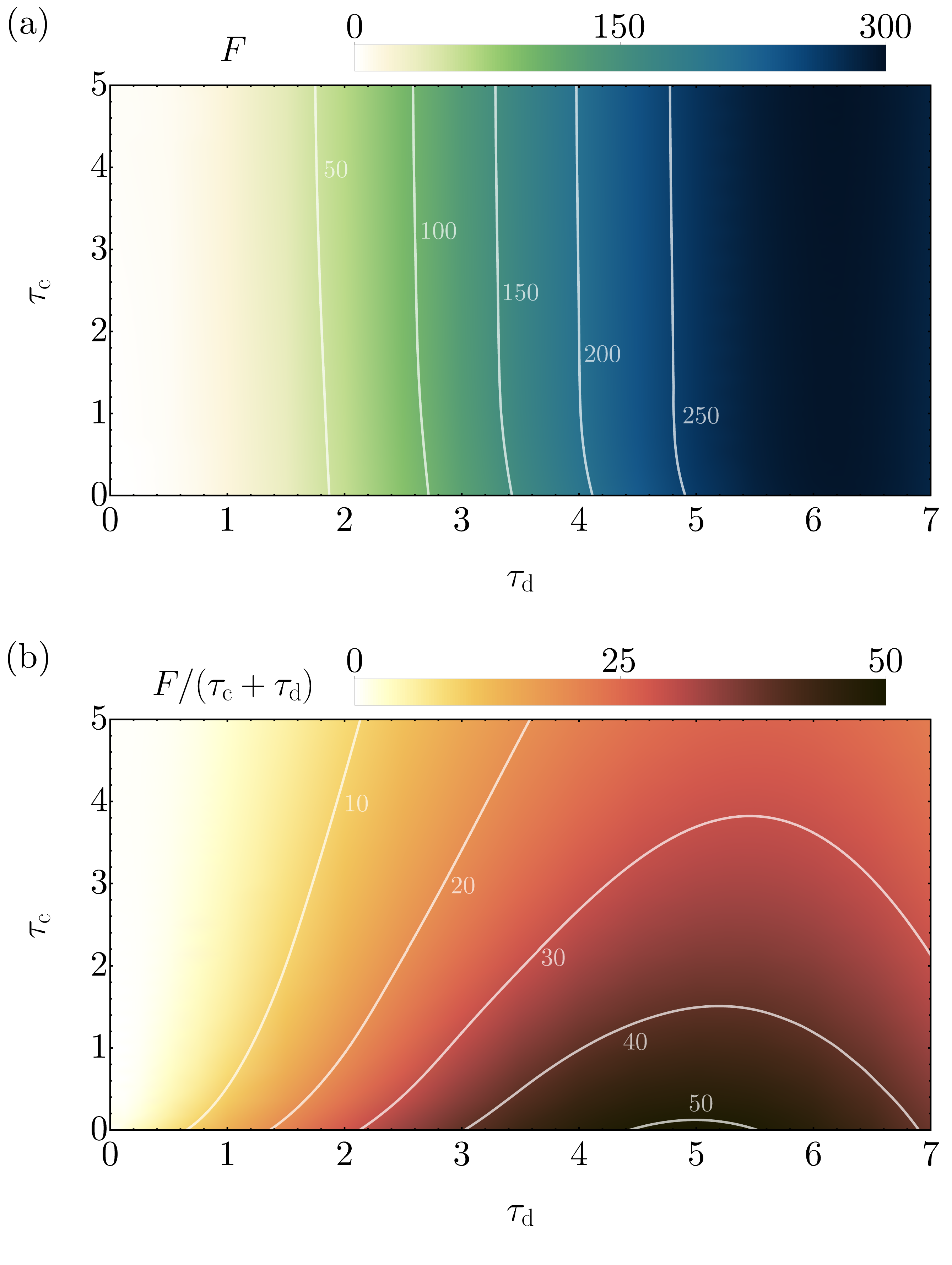}
\caption{Optimization of cooling and driving times for trapped-ion hot-state sensing using full numerical master-equation simulations.
(a) Full QFI as a function of cooling time $\CoolTime$ and driving time $\CatTime$.
(b) Effective metrological gain per unit preparation time, $\QFisher/(\CoolTime+\CatTime)$, showing the emergence of an optimal finite driving time despite the monotonic increase of $\QFisher$.
The calculations assume initial thermal occupation $\ThermalOccupation=10$, cooling rate $\CoolingRate=1$, motional-heating rate $\HeatingRate=10^{-3}$, and ancilla-spin-dephasing rate $\SpinDephasingRate=10^{-2}$, corresponding to experimentally realistic trapped-ion parameters.
The optimal driving time is set by the competition between coherent QFI growth and noise-induced degradation during state preparation.
\label{Fig2}}
\end{figure}

Fig.~\ref{Fig2} presents full master-equation simulations for experimentally realistic trapped-ion parameters and reproduces the same qualitative behavior.
A clear optimum emerges at finite preparation time, whereas the optimal cooling duration remains close to zero.
Although additional cooling slightly improves the single-shot QFI, the gain is insufficient to compensate for the corresponding increase in protocol duration.
For the parameters of Fig.~\ref{Fig2}, the optimum is quantitatively consistent with the heating-dominated estimate $\Time_{\mathrm{d}}^\star=5$ obtained for $\HeatingRate=10^{-3}$.
The conclusion is not that cooling is universally detrimental.
Cooling can become advantageous when the local thermal contribution is comparable to the branch contribution, when thermal occupation modifies subsequent decoherence or readout, or when the large-separation approximation breaks down.
Within the effective model considered here, however, optimal displacement sensing is achieved not by maximizing ground-state purity, but by maximizing branch coherence per unit total experimental time.

\textit{Conclusions and outlook}---We established hot-state engineering as a route toward quantum-enhanced bosonic sensing beyond the standard paradigm based on near-ground-state initialization.
Thermal occupation does not universally destroy displacement sensitivity.
The crucial question is how the preparation protocol structures the thermal ensemble and where the metrological resource is encoded.
Parity engineering restores large QFI through Hilbert-space exclusion between opposite parity sectors, whereas coherent branch engineering allows catlike sensitivity to survive directly within thermally occupied states.
The same structures determine the dominant decoherence mechanisms: parity-engineered protocols are limited by parity-breaking noise, while branch-engineered cats are limited by decoherence between displaced branches.
Thus, quantum-enhanced sensing does not require purification of the full oscillator state, but protection or structuring of the degree of freedom carrying the displacement-sensitive coherence.
This viewpoint is conceptually related to noiseless-subsystem, decoherence-free-subspace, and reference-frame ideas, where useful quantum information is encoded in selected symmetry or relational degrees of freedom rather than in the full physical Hilbert space~\cite{Zanardi1997,Lidar1998,Knill2000,Lidar2003,Bartlett2007}.

Our findings are relevant for trapped ions~\cite{Millican2025,Matsos2024,Valahu2025}, bulk acoustic~\cite{vonLupke2024,Rahman2025} and optomechanical~\cite{Mercade2023,Madiot2023} resonators, circuit quantum electrodynamics devices~\cite{Chakram2021,Heeres2017,Krisnanda2025,Pan2025}, and levitated mechanical systems~\cite{Gonzalez-Ballestero2021}.
More broadly, the optimal sensing strategy is determined not only by maximal single-shot QFI, but by the interplay between cooling overhead, state preparation, decoherence, and repetition rate.
Within the effective ECD model, this optimum is obtained by maximizing coherent branch generation per unit total experimental time rather than ground-state purity.
Future directions include extending hot-state metrology to phase-insensitive~\cite{Grochowski2025} and distributed schemes~\cite{Grochowski2026}, exploring hot-state preparation through Hamiltonian or potential engineering~\cite{Casulleras2024,Grochowski2025a,Grochowski2026a}, and investigating analogous mechanisms in strongly nonlinear or many-body bosonic systems.

\begin{acknowledgments}
\textit{Acknowledgments}---We thank R. Filip for useful discussions.
P.T.G. was supported by the project CZ.02.01.01/00/22\_010/0013054 (C-MONS) within Programme JAC MSCA Fellowships at Palacký University Olomouc IV.
We acknowledge the open-source package QuTiP, used for numerical simulations~\cite{Johansson2012,Johansson2013,Lambert2024}.
Data analysis and simulation codes will be available on Zenodo before publication~\cite{zenodo4}.
Some plots were produced using Scientific colour maps~\cite{Crameri2020,Crameri2023}.
\end{acknowledgments}

\bibliography{HotCats}

\clearpage
\part*{}
\vspace{-4em}

\title{Supplemental Material: To Cool, or Not to Cool? Displacement Sensing with Hot Quantum States}

\maketitle

\makeatletter
\renewcommand{\c@secnumdepth}{0}
\makeatother

\renewcommand{\thesection}{S\Roman{section}}
\renewcommand{\thesubsection}{\thesection.\arabic{subsection}}
\renewcommand{\thesubsubsection}{\thesubsection.\arabic{subsubsection}}

\renewcommand{\thefigure}{S\arabic{figure}}
\renewcommand{\theequation}{S\arabic{equation}}

\makeatletter
\renewcommand{\p@subsection}{}
\renewcommand{\p@subsubsection}{}
\renewcommand{\p@figure}{}
\makeatother

\setcounter{section}{0}
\setcounter{equation}{0}
\setcounter{figure}{0}

\onecolumngrid
\smtableofcontents
\vspace{1cm}

\newpage 
\twocolumngrid

\smsection{Squeezed thermal states}
Let us start with the definition of the thermal state,
\begin{align}
    \ThermalState
    =
    \rounds{1-\ResPar}
    \sum_{n=0}^{\infty}
    \ResPar^n
    \ket{n}\bra{n},
    \quad
    \ResPar
    =
    \frac{
    \ThermalOccupation
    }{
    1+\ThermalOccupation
    }.
    \label{eq:thermal_state}
\end{align}
For any state diagonal in the Fock basis,
\begin{align}
    \DensityMatrix
    =
    \sum_n \Prob_n\ket{n}\bra{n},
\end{align}
Eq.~\eqref{eq:qfi_general} gives
\begin{align}
    \QFisher
    =
    4
    \sum_{n=0}^{\infty}
    \rounds{n+1}
    \frac{
    \rounds{\Prob_n-\Prob_{n+1}}^2
    }{
    \Prob_n+\Prob_{n+1}
    }.
    \label{eq:qfi_diagonal}
\end{align}
For the thermal state, $\Prob_{n+1}=\ResPar \Prob_n$, hence
\begin{align}
    \QFisher_\ThLetter
    &=
    4
    \frac{
    \rounds{1-\ResPar}^2
    }{
    1+\ResPar
    }
    \sum_{n=0}^{\infty}
    \rounds{n+1}p_n
    =
    4
    \frac{
    \rounds{1-\ResPar}^2
    }{
    1+\ResPar
    }
    \rounds{\ThermalOccupation+1}
    \nonumber\\
    &=
    4
    \frac{
    1-\ResPar
    }{
    1+\ResPar
    }
    =
    \frac{
    4
    }{
    2\ThermalOccupation+1
    }.
    \label{eq:qfi_thermal}
\end{align}
Let us define the squeezed thermal state as
\begin{align}
    \SqueezedThermalState
    =
    \SqueezingOp{-\Squeeze}
    \ThermalState
    \SqueezingOpDagger{-\Squeeze}
\end{align}
with the squeezing parameter $\Squeeze$ and the squeezing operator $ \SqueezingOp{\cdot}$.
The sign is chosen such that the sensed quadrature is anti-squeezed,
\begin{align}
    \SqueezingOpDagger{-\Squeeze}
    \Quadrature
    \SqueezingOp{-\Squeeze}
    =
    e^{\Squeeze}\Quadrature .
    \label{eq:squeezing}
\end{align}
Using unitary covariance of the QFI,
\begin{align}
    \QFisher\squares{\SqueezedThermalState,\Quadrature}
    &=
    \QFisher\squares{
    \ThermalState,
    \SqueezingOpDagger{-\Squeeze}
    \Quadrature
    \SqueezingOp{-\Squeeze}
    }
    =
    \QFisher\squares{
    \ThermalState,
    e^{\Squeeze}\Quadrature
    }.
\end{align}
The QFI is quadratic in the generator, and therefore
\begin{align}
    \QFisher_{\text{sq}}
    =
    e^{2\Squeeze}
    \QFisher_\ThLetter
    =
    \frac{
    4e^{2\Squeeze}
    }{
    2\ThermalOccupation+1
    }.
    \label{eq:qfi_squeezed}
\end{align}
The total mean occupation is obtained from
\begin{align}
    \SqueezingOpDagger{\Squeeze}
    \Annihilation
    \SqueezingOp{\Squeeze}
    =
    \Annihilation \cosh \Squeeze
    +
    \Creation \sinh \Squeeze .
\end{align}
Hence
\begin{align}
    \angles{\NumberOperator}_{\text{sq}}
    &=
    \ThermalOccupation \cosh^2\Squeeze
    +
    \rounds{\ThermalOccupation+1}\sinh^2\Squeeze
    \nonumber\\
    &=
    \ThermalOccupation
    +
    \rounds{2\ThermalOccupation+1}
    \sinh^2\Squeeze .
    \label{eq:ntot_squeezed}
\end{align}
This shows the enhanced energy cost of squeezing a thermal state due to bosonic statistics.
The exemplary state is shown in Fig.~\ref{FigSWigners}(a).

\smsubsection{Parity-filtered squeezed thermal states}
Let us now consider the effect of parity filtering prior to squeezing.
Starting from a bosonic state $\DensityMatrix$, parity projection produces
\begin{align}
    \DensityMatrix_{\pm}
    =
    \frac{
    \Projector_{\pm}
    \DensityMatrix
    \Projector_{\pm}
    }{
    \Trace{
    \Projector_{\pm}
    \DensityMatrix
    }
    }.
\end{align}
Applying squeezing then gives
\begin{align}
    \DensityMatrix_{\SqLetter,\pm}
    =
    \SqueezingOperator(-\Squeeze)
    \DensityMatrix_{\pm}
    \SqueezingOperator^\dagger(-\SqueezingParameter).
\end{align}
Since squeezing preserves parity,
\begin{align}
    \commutator{
    \SqueezingOperator(-\Squeeze)
    }{
    \ParityOp
    }
    =
    0,
\end{align}
the squeezed state remains entirely confined to a single parity sector,
\begin{align}
    \DensityMatrix_{\SqLetter,\pm}
    =
    \Projector_{\pm}
    \DensityMatrix_{\SqLetter,\pm}
    \Projector_{\pm}.
\end{align}
Consequently, as also argued in the main text, the QFI simplifies exactly to
\begin{align}
    \QFisher_{\SqLetter,\pm}
    =
    4
    \Trace{
    \DensityMatrix_{\SqLetter,\pm}
    \Quadrature^2
    }.
\end{align}
Using the squeezing transformation~\eqref{eq:squeezing} for the amplified quadrature, we obtain
\begin{align}
    \QFisher_{\SqLetter,\pm}
    =
    4
    e^{2\Squeeze}
    \Trace{
    \DensityMatrix_{\pm}
    \Quadrature^2
    }.
\end{align}
For parity-filtered states diagonal in the Fock basis,
\begin{align}
    \angles{
    \Annihilation^2
    }_{\DensityMatrix_\pm}
    =
    0,
\end{align}
which gives
\begin{align}
    \QFisher_{\SqLetter,\pm}
    =
    4
    e^{2\Squeeze}
    \rounds{
    2
    \angles{
    \Creation\Annihilation
    }_{\DensityMatrix_\pm}
    +1
    }.
    \label{eq:qfi_parity_filtered_squeezing}
\end{align}
Equation~\eqref{eq:qfi_parity_filtered_squeezing} shows that parity filtering qualitatively changes the role of thermal occupation.
Without parity projection, thermal fluctuations suppress the squeezing-enhanced displacement sensitivity.
After parity filtering, however, the displacement generator couples the occupied parity ladder to an initially empty orthogonal sector.
As a result, increasing occupation of the populated parity manifold enhances the susceptibility rather than degrading it.
For an even-parity thermal state,
\begin{align}
    \ThermalState^{+}=
    \rounds{
    1-\ResPar^2
    }
    \sum_{m=0}^{\infty}
    \ResPar^{2m} \ket{2m} \bra{2m},
\end{align}
the mean occupation becomes
\begin{align}
    \angles{
    \Creation\Annihilation
    }_{+}
    =
    \frac{
    2\ThermalOccupation^2
    }{
    2\ThermalOccupation+1
    }.
\end{align}
Substituting into Eq.~\eqref{eq:qfi_parity_filtered_squeezing} gives
\begin{align}
    \QFisher_{\mathrm{sq},+}
    =
    4
    e^{2\SqueezingParameter}
    \rounds{
    \frac{
    4\ThermalOccupation^2
    }{
    2\ThermalOccupation+1
    }
    +1
    }.
\end{align}
For large thermal occupation, $\ThermalOccupation\gg1$, this scales asymptotically as
\begin{align}
    \QFisher_{\mathrm{sq},+}
    \simeq
    8
    e^{2\SqueezingParameter}
    \ThermalOccupation .
\end{align}
Thus, parity filtering completely reverses the thermal scaling of squeezed-state displacement sensing.
Instead of suppressing the metrological enhancement, thermal occupation becomes a resource that amplifies the displacement susceptibility within the populated parity sector.
The exemplary state is shown in Fig.~\ref{FigSWigners}(b).

\smsection{Hot Fock states}

\begin{figure}[ht!]
    \includegraphics[width=\linewidth]{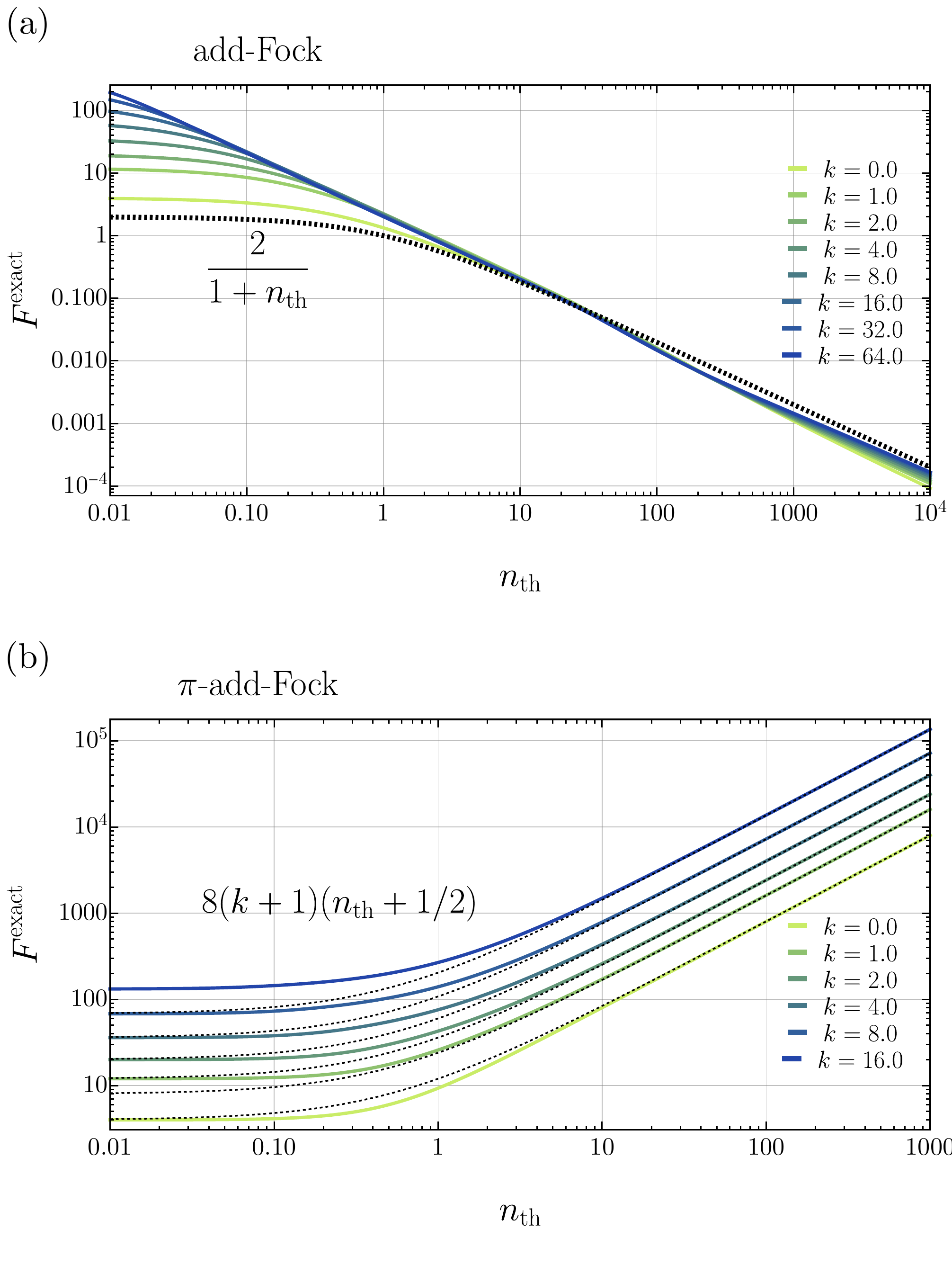}  
    \caption{Quantum Fisher information for displacement sensing using phonon-added hot-Fock states. 
(a) Standard phonon-added hot-Fock states generated by applying $k$ creation operators to a thermal state before cat-state synthesis. 
The QFI decreases asymptotically independently of the number of additions, showing that phonon addition alone does not remove the thermal mixed-state suppression. 
(b) Parity-filtered phonon-added hot-Fock states. 
After parity projection, the QFI exhibits the enhancement, demonstrating that parity filtering converts phonon addition into an energy-enhanced metrological resource. 
Solid curves show exact numerical results, while dotted black curves denote the corresponding asymptotic analytical estimates.
  \label{FigSFock}}
\end{figure}

\smsubsection{Phonon addition}
Let us define a $k$-phonon-added thermal state as
\begin{align}
    \HotFockState_{\AddLetter}^{(k)}
    =
    \frac{
    \Annihilation^{\dagger k}
    \ThermalState
    \Annihilation^k
    }{
    \Tr\rounds{
    \Annihilation^{\dagger k}
    \ThermalState
    \Annihilation^k
    }
    },
    \label{eq:phonon_added_def}
\end{align}
where $k$ is the number of applied creation operators. 
Note that the mean occupation increase is not equal to $k$ for a thermal input.
This is because the map is nonunitary and postselected.
Using
\begin{align}
    \Annihilation^{\dagger k}\ket{m}
    =
    \sqrt{
    \frac{
    \rounds{m+k}!
    }{
    m!
    }
    }
    \ket{m+k},
\end{align}
we have
\begin{align}
    \Annihilation^{\dagger k}
    \ThermalState
    \Annihilation^k
    =
    \rounds{1-\ResPar}
    \sum_{m=0}^{\infty}
    \ResPar^m
    \frac{
    \rounds{m+k}!
    }{
    m!
    }
    \ket{m+k}\bra{m+k}.
\end{align}
The normalization is
\begin{align}
    \Tr\rounds{
    \Annihilation^{\dagger k}
    \ThermalState
    \Annihilation^k
    }
    &=
    \rounds{1-\ResPar}
    \sum_{m=0}^{\infty}
    \ResPar^m
    \frac{
    \rounds{m+k}!
    }{
    m!
    }
    \nonumber\\
    &=
    k!
    \rounds{1-\ResPar}
    \sum_{m=0}^{\infty}
    \binom{m+k}{k}
    \ResPar^m
    =
    \frac{
    k!
    }{
    \rounds{1-\ResPar}^{k}
    }.
\end{align}
Thus,
\begin{align}
    \HotFockState_{\AddLetter}^{(k)}
    =
    \sum_{n=k}^{\infty}
    p_{n,\AddLetter}^{(k)}
    \ket{n}\bra{n},
\end{align}
with
\begin{align}
    \Prob_{n,\AddLetter}^{(k)}
    =
    \binom{n}{k}
    \rounds{1-\ResPar}^{k+1}
    \ResPar^{n-k},
    \quad
    n\geq k,
    \label{eq:p_phonon_added}
\end{align}
and $p_{n,\AddLetter}^{(k)}=0$ for $n<k$.
This is a negative-binomial distribution.
The mean occupation is
\begin{align}
    \angles{\Creation \Annihilation}_{\AddLetter}
    &=
    \sum_{n=k}^{\infty}
    n p_{n,\AddLetter}^{(k)}
    =
    k+\rounds{k+1}\ThermalOccupation.
    \label{eq:mean_phonon_added}
\end{align}
Therefore
\begin{align}
    \angles{\NumberOperator}_{\AddLetter}
    -
    \ThermalOccupation
    =
    k\rounds{1+\ThermalOccupation}.
\end{align}
The number of applied additions is $k$, but the mean energy increase is Bose-enhanced.
Since the state is diagonal in Fock basis, the exact QFI is
\begin{align}
    \QFisher_{\AddLetter}^{(k)}
    =
    4
    \sum_{n=0}^{\infty}
    \rounds{n+1}
    \frac{
    \rounds{
    p_{n,\AddLetter}^{(k)}
    -
    p_{n+1,\AddLetter}^{(k)}
    }^2
    }{
    p_{n,\AddLetter}^{(k)}
    +
    p_{n+1,\AddLetter}^{(k)}
    }.
    \label{eq:qfi_phonon_added}
\end{align}
It is useful to extract the large-temperature scaling of Eq.~\eqref{eq:qfi_phonon_added}.
Let us define an auxiliary variable,
\begin{align}
    \epsilon
    =
    1-\ResPar
    =
    \frac{1}{1+\ThermalOccupation}.
\end{align}
Writing $n=m+k$, the ratio of adjacent probabilities is
\begin{align}
    r_m
    =
    \frac{
    p_{m+k+1,\AddLetter}^{(k)}
    }{
    p_{m+k,\AddLetter}^{(k)}
    }
    =
    \ResPar
    \frac{
    m+k+1
    }{
    m+1
    }.
\end{align}
The QFI can therefore be written as
\begin{align}
    &\QFisher_{\AddLetter}^{(k)}
    = \nonumber \\
    &4k
    \rounds{1-\ResPar}^{k+1}
    +
    4
    \sum_{m=0}^{\infty}
    \rounds{m+k+1}
    p_{m+k,\AddLetter}^{(k)}
    \frac{
    \rounds{1-r_m}^2
    }{
    1+r_m
    }.
    \label{eq:qfi_phonon_added_ratio}
\end{align}
For $\ThermalOccupation\gg1$, the relevant values of $m$ scale as
\begin{align}
    x
    =
    \epsilon m
    =
    \mathcal O(1),
\end{align}
where $x$ is another auxiliary variable.
The negative-binomial distribution then approaches the Gamma distribution
\begin{align}
    p_{m+k,\AddLetter}^{(k)}
    \simeq
    \epsilon
    \frac{
    x^k e^{-x}
    }{
    k!
    },
\end{align}
while
\begin{align}
    1-r_m
    =
    \epsilon
    \rounds{
    1-\frac{k}{x}
    }
    +
    \mathcal O\rounds{\epsilon^2}.
\end{align}
Substituting these expressions into Eq.~\eqref{eq:qfi_phonon_added_ratio} gives
\begin{align}
    \QFisher_{\AddLetter}^{(k)}
    &\simeq
    2\epsilon
    \frac{1}{k!}
    \int_0^\infty
    dx\,
    e^{-x}
    x^{k+1}
    \rounds{
    1-\frac{k}{x}
    }^2
    \nonumber\\
    &=
    2\epsilon
    =
    \frac{
    2
    }{
    1+\ThermalOccupation
    }.
    \label{eq:qfi_phonon_added_asymptotic}
\end{align}
Thus, at fixed $k$ and large $\ThermalOccupation$,
\begin{align}
    \QFisher_{\AddLetter}^{(k)}
    \sim
    \frac{2}{\ThermalOccupation}.
\end{align}
The leading asymptotic scaling is independent of $k$.
This shows that phonon addition strongly increases the mean energy but does not remove the mixed-state spectral suppression caused by the simultaneous occupation of neighboring Fock sectors.
The comparison between the exact values and the asymptotic expression is plotted in Fig.~\ref{FigSFock}(a).
For $\ThermalOccupation=0$, only $p_{k,\AddLetter}^{(k)}=1$ is nonzero, and the state is $\ket{k}$. Then
\begin{align}
    \QFisher_{\AddLetter}^{(k)}
    =
    4
    \bra{k}
    \Quadrature^2
    \ket{k}
    =
    4\rounds{2k+1}.
\end{align}
The exemplary state is shown in Fig.~\ref{FigSWigners}(c).

\smsubsubsection{Parity-filtered phonon addition}
We now consider phonon addition after parity filtering.
Starting from the parity-projected state
\begin{align}
    \DensityMatrix_{\pm}
    =
    \frac{
    \Projector_{\pm}
    \DensityMatrix
    \Projector_{\pm}
    }{
    \Tr(
    \Projector_{\pm}
    \DensityMatrix
    \Projector_{\pm})
    }
    ,
\end{align}
we apply $k$ phonon additions,
\begin{align}
    \DensityMatrix^{(k),\pm}_{\AddLetter}
    =
    \frac{
    \Annihilation^{\dagger k}
    \DensityMatrix_{\pm}
    \Annihilation^k
    }{
    \Trace(
    \Annihilation^{\dagger k}
    \DensityMatrix_{\pm}
    \Annihilation^k
    )
    }.
\end{align}
Phonon addition changes parity depending on $k$. Since
\begin{align}
    \ParityOp \Annihilation^{\dagger k} \ParityOp
    =
    \rounds{-1}^k
    \Annihilation^{\dagger k},
\end{align}
the final state still occupies a single parity sector.
It is in the same parity sector for even $k$ and in the opposite parity sector for odd $k$.
Therefore, the general parity-projected QFI identity applies,
\begin{align}
    \QFisher^{(k),\pm}_{\AddLetter}
    =
    4
    \Trace[
    \DensityMatrix^{(k),\pm}_{\AddLetter}
    \rounds{
    \Creation+\Annihilation
    }^2
    ],
\end{align}
or equivalently,
\begin{align}
    \QFisher^{(k),\pm}_{\AddLetter}
    =
    4
    \rounds{
    2
    \angles{
    \Creation\Annihilation
    }_{(k),\pm}
    +1
    +
    2\mathrm{Re}
    \angles{
    \Annihilation^2
    }_{(k),\pm}
    }.
\end{align}
For parity-filtered thermal states, the density matrix is diagonal in the Fock basis before and after phonon addition.
Hence,
\begin{align}
    \angles{
    \Annihilation^2
    }_{k,\pm}
    =
    0,
\end{align}
and
\begin{align}
    \QFisher^{(k),\pm}_{\AddLetter}
    =
    4
    \rounds{
    2
    \angles{
    \Creation\Annihilation
    }_{k,\pm}
    +1
    }.
    \label{eq:qfi_parity_filtered_phonon_addition}
\end{align}
For the even-filtered thermal state,
\begin{align}
    \ThermalState^{+}
    =
    \rounds{1-\ResPar^2}
    \sum_{m=0}^{\infty}
    \ResPar^{2m}
    \ket{2m}\bra{2m},
\end{align}
the $k$-phonon-added state has weights proportional to $\frac{\rounds{2m+k}!}{
    \rounds{2m}!
    }
    \ResPar^{2m}$.
Thus,
\begin{align}
    \DensityMatrix^{(k),+}_{\AddLetter}
    =
    \frac{
    \sum_{m=0}^{\infty}
    \frac{
    \rounds{2m+k}!
    }{
    \rounds{2m}!
    }
    \ResPar^{2m}
    \ket{2m+k}\bra{2m+k}
    }{
    \sum_{m=0}^{\infty}
    \frac{
    \rounds{2m+k}!
    }{
    \rounds{2m}!
    }
    \ResPar^{2m}
    }.
\end{align}
The mean occupation is therefore
\begin{align}
    \angles{
    \Creation\Annihilation
    }_{k,+}
    =
    k
    +
    \frac{
    \sum_{m=0}^{\infty}
    2m
    \frac{
    \rounds{2m+k}!
    }{
    \rounds{2m}!
    }
    \ResPar^{2m}
    }{
    \sum_{m=0}^{\infty}
    \frac{
    \rounds{2m+k}!
    }{
    \rounds{2m}!
    }
    \ResPar^{2m}
    }.
\end{align}
Substitution into Eq.~\eqref{eq:qfi_parity_filtered_phonon_addition} gives the exact QFI for parity-filtered phonon addition.
The important point is structural.
Without parity projection, phonon addition increases the mean occupation but does not empty the opposite parity sector.
The mixed-state QFI, therefore, remains suppressed by population overlap between adjacent Fock states.
With parity filtering, one parity ladder is removed before phonon addition.
Thus, parity filtering changes phonon addition from an energy-increasing operation into a parity-structured metrological resource.
It is further useful to derive the asymptotic scaling of the parity-filtered phonon-added state.
An explicit analytical estimate can be obtained from the normalization sum.
Let us define
\begin{align}
    Z_k^{+}
    =
    \sum_{m=0}^{\infty}
    \frac{
    \rounds{2m+k}!
    }{
    \rounds{2m}!
    }
    \ResPar^{2m}.
\end{align}
Then the mean occupation of the parity-filtered, $k$-phonon-added state can be written as
\begin{align}
    \angles{
    \NumberOperator
    }_{k,+}
    =
    k
    +
    \ResPar
    \partial_{\ResPar}
    \log Z_k^{+}.
    \label{eq:mean_parity_added_generator}
\end{align}
The sum $Z_k^{+}$ admits the useful representation
\begin{align}
    Z_k^{+}
    =
    \partial_{\ResPar}^{k}
    \frac{
    \ResPar^k
    }{
    1-\ResPar^2
    },
    \label{eq:Zk_even_added}
\end{align}
because
\begin{align}
    \partial_{\ResPar}^{k}
    \ResPar^{2m+k}
    =
    \frac{
    \rounds{2m+k}!
    }{
    \rounds{2m}!
    }
    \ResPar^{2m}.
\end{align}
Equations~\eqref{eq:mean_parity_added_generator} and~\eqref{eq:Zk_even_added} give an exact compact form for the QFI,
\begin{align}
    \QFisher^{(k),+}_{\AddLetter}
    =
    4
    \rounds{
    2k
    +
    2\ResPar
    \partial_{\ResPar}
    \log Z_k^{+}
    +
    1
    }.
    \label{eq:qfi_parity_added_exact_generator}
\end{align}

The scaling follows immediately from the singular behavior of $Z_k^{+}$ at high temperature.
Writing again
\begin{align}
    \epsilon
    =
    1-\ResPar
    =
    \frac{
    1
    }{
    1+\ThermalOccupation
    },
\end{align}
we have, for fixed $k$ and $\epsilon\ll1$,
\begin{align}
    Z_k^{+}
    =
    \partial_{\ResPar}^{k}
    \frac{
    \ResPar^k
    }{
    1-\ResPar^2
    }
    \simeq
    \frac{
    k!
    }{
    2
    \epsilon^{k+1}
    }.
\end{align}
Therefore,
\begin{align}
    \ResPar
    \partial_{\ResPar}
    \log Z_k^{+}
    =
    \rounds{k+1}
    \frac{
    \ResPar
    }{
    1-\ResPar
    }
    +
    \mathcal O(1)
    =
    \rounds{k+1}
    \ThermalOccupation
    +
    \mathcal O(1).
\end{align}
Substitution into Eq.~\eqref{eq:qfi_parity_added_exact_generator} gives
\begin{align}
    \QFisher^{(k),+}_{\AddLetter}
    =
    8 \rounds{k+1} (\ThermalOccupation + 1/2) + 4+ \mathcal O(1)
    .
    \label{eq:qfi_parity_added_asymptotic}
\end{align}
Thus, at fixed $k$, parity-filtered phonon addition produces a QFI that grows linearly with the initial thermal occupation, with a prefactor proportional to $k+1$.
In Fig.~\ref{FigSFock}(b), we provide a numerical check of this approximation.
The exemplary state is shown in Fig.~\ref{FigSWigners}(d).

\smsubsection{Jaynes--Cummings ladder climbing}
We now consider deterministic Fock-state synthesis using a sequence of resonant Jaynes--Cummings (JC) interactions. 
The goal is to construct a pulse sequence that maps
\begin{align}
    \ket{0}
    \rightarrow
    \ket{1}
    \rightarrow
    \cdots
    \rightarrow
    \ket{k},
\end{align}
generating highly excited Fock states from the oscillator ground state.
The resonant JC Hamiltonian in the rotating frame is
\begin{align}
    \Hamiltonian_\JCLetter
    =
    \hbar \JCCoup
    \rounds{
    \Annihilation \sigmaUpOp
    +
    \Creation \sigmaDownOp
    },
\end{align}
where $\sigmaUpOp=\ket{e}\bra{g}$ and $\sigmaDownOp=\ket{g}\bra{e}$ are qubit raising and lowering operators, respectively.
The Hamiltonian conserves the total excitation number,
\begin{align}
    \commutator{
    \Hamiltonian_\JCLetter
    }{
    \Creation\Annihilation
    +
    \sigmaUpOp\sigmaDownOp
    }
    =
    0,
\end{align}
and therefore decomposes into independent two-dimensional manifolds
\begin{align}
    \curlies{
    \ket{e,n},
    \ket{g,n+1}
    }.
\end{align}
Within each manifold, the exact evolution is
\begin{align}
    \ket{e,n}
    \rightarrow
    \cos\rounds{
    \JCCoup
    \sqrt{n+1}
    \DimTime
    }
    \ket{e,n}
    -
    \ImagUnit
    \sin\rounds{
    \JCCoup
    \sqrt{n+1}
    \DimTime
    }
    \ket{g,n+1}.
    \label{eq:jc_evolution}
\end{align}
To synthesize $\ket{k}$ from vacuum, we use a calibrated pulse sequence.
At step $j=0,\ldots,k-1$, the qubit is prepared in $\ket{e}$ and the pulse duration is chosen as
\begin{align}
    \DimTime_j
    =
    \frac{
    \pi
    }{
    2\JCCoup\sqrt{j+1}
    }.
    \label{eq:jc_calibrated_pulse}
\end{align}
This choice produces a perfect $\pi$-pulse in the manifold $\curlies{ \ket{e,j}, \ket{g,j+1}}$ with
\begin{align}
    \ket{e,j}
    \rightarrow
    -\ImagUnit
    \ket{g,j+1}.
\end{align}
After resetting the qubit and repeating the procedure, the protocol deterministically generates
\begin{align}
    \ket{0}
    \rightarrow
    \ket{1}
    \rightarrow
    \cdots
    \rightarrow
    \ket{k}.
\end{align}
The important point is that the same pulse sequence acts differently on higher Fock states.
For a general initial oscillator state $\ket{n}$, the $j$th pulse produces
\begin{align}
    \ket{e,n}
    \rightarrow
    \sqrt{
    1-s_n^{(j)}
    }
    \ket{e,n}
    -
    \ImagUnit
    \sqrt{
    s_n^{(j)}
    }
    \ket{g,n+1},
\end{align}
where
\begin{align}
    s_n^{(j)}
    =
    \sin^2\rounds{
    \frac{\pi}{2}
    \sqrt{
    \frac{n+1}{j+1}
    }
    }.
    \label{eq:jc_transfer_probability}
\end{align}
After tracing out or resetting the qubit, the oscillator component $\ket{n}\bra{n}$ evolves into
\begin{align}
    \ket{n}\bra{n}
    \rightarrow
    \rounds{
    1-s_n^{(j)}
    }
    \ket{n}\bra{n}
    +
    s_n^{(j)}
    \ket{n+1}\bra{n+1}.
    \label{eq:nn_evolution}
\end{align}
Then, a single JC pulse only couples neighboring Fock sectors.
After multiple calibrated pulses, however, repeated branching generates a distribution over several shifted number states.
Let us define $P_k(\ell|n)$ as the probability that an initial $\ket{n}$ has climbed upward by $\ell$ excitations after $k$ pulses.
Then
\begin{align}
    \ket{n}\bra{n}
    \rightarrow
    \rho_n^{(k)}
    =
    \sum_{\ell=0}^{k}
    P_k(\ell|n)
    \ket{n+\ell}\bra{n+\ell}.
    \label{eq:jc_kernel_state}
\end{align}
The transition probabilities satisfy the recursion
\begin{align}
    P_{j+1}(\ell|n)
    &=
    P_j(\ell|n)
    \rounds{
    1-s_{n+\ell}^{(j)}
    }
+
    P_j(\ell-1|n)
    s_{n+\ell-1}^{(j)},
    \label{eq:jc_kernel_recursion}
\end{align}
with $ P_0(\ell|n) = \delta_{\ell,0}$.
Eq.~\eqref{eq:jc_kernel_recursion} has the exact path-sum solution
\begin{align}
    P_k(\ell|n)
    =
    \sum_{\substack{
    x_0,\ldots,x_{k-1}\in\{0,1\}\\
    \sum_j x_j=\ell
    }}
    \prod_{j=0}^{k-1}
    \rounds{
    s_{n+X_j}^{(j)}
    }^{x_j}
    \rounds{
    1-s_{n+X_j}^{(j)}
    }^{1-x_j},
    \label{eq:jc_path_sum}
\end{align}
where
\begin{align}
    X_j
    =
    \sum_{r=0}^{j-1}
    x_r
\end{align}
is the net increase of the oscillator occupation number before pulse $j$.
Therefore, the oscillator occupation entering pulse $j$ is $n+X_j$.
For $n=0$, the calibrated sequence produces a perfect state-to-state transfer.
Indeed,
\begin{align}
    s_j^{(j)}
    =
    1,
\end{align}
at every step of the ladder, so the vacuum component follows the unique path
\begin{align}
    \ket{0}
    \rightarrow
    \ket{1}
    \rightarrow
    \cdots
    \rightarrow
    \ket{k}.
\end{align}
Equivalently,
\begin{align}
    P_k(\ell|0)
    =
    \delta_{\ell,k},
\end{align}
and the protocol exactly prepares $\ket{k}$ from vacuum.
For $n>0$, however, the transfer probabilities are generally neither zero nor one.
The same calibrated pulse sequence, therefore, generates a mixed distribution over
\begin{align}
    \ket{n},
    \ket{n+1},
    \ldots,
    \ket{n+k}.
\end{align}
This is the fundamental origin of the desynchronization of JC ladder climbing on thermal states.
For a thermal input state,
\begin{align}
    \ThermalState
    =
    \sum_n
    p_n
    \ket{n}\bra{n},
\end{align}
each Fock component evolves independently according to Eq.~\eqref{eq:jc_kernel_state}. 
The final oscillator state is therefore
\begin{align}
    \HotFockState_{\JCLetter}^{(k)}
    &=
    \sum_n
    p_n
    \rho_n^{(k)}
   =
    \sum_{m=0}^{\infty}
    p_m^{(k)}
    \ket{m}\bra{m},
\end{align}
with
\begin{align}
    p_m^{(k)}
    =
    \sum_n
    p_n
    P_k(m-n|n).
    \label{eq:jc_final_distribution}
\end{align}
Since the final state remains diagonal in the Fock basis, the exact displacement QFI is
\begin{align}
    \QFisher_{\JCLetter}^{(k)}
    =
    4
    \sum_{m=0}^{\infty}
    \rounds{m+1}
    \frac{
    \rounds{
    p_m^{(k)}
    -
    p_{m+1}^{(k)}
    }^2
    }{
    p_m^{(k)}
    +
    p_{m+1}^{(k)}
    },
    \label{eq:qfi_jc_final}
\end{align}
which can be extracted numerically.
The exemplary state is shown in Fig.~\ref{FigSWigners}(e).

\begin{figure}[ht!]
    \includegraphics[width=\linewidth]{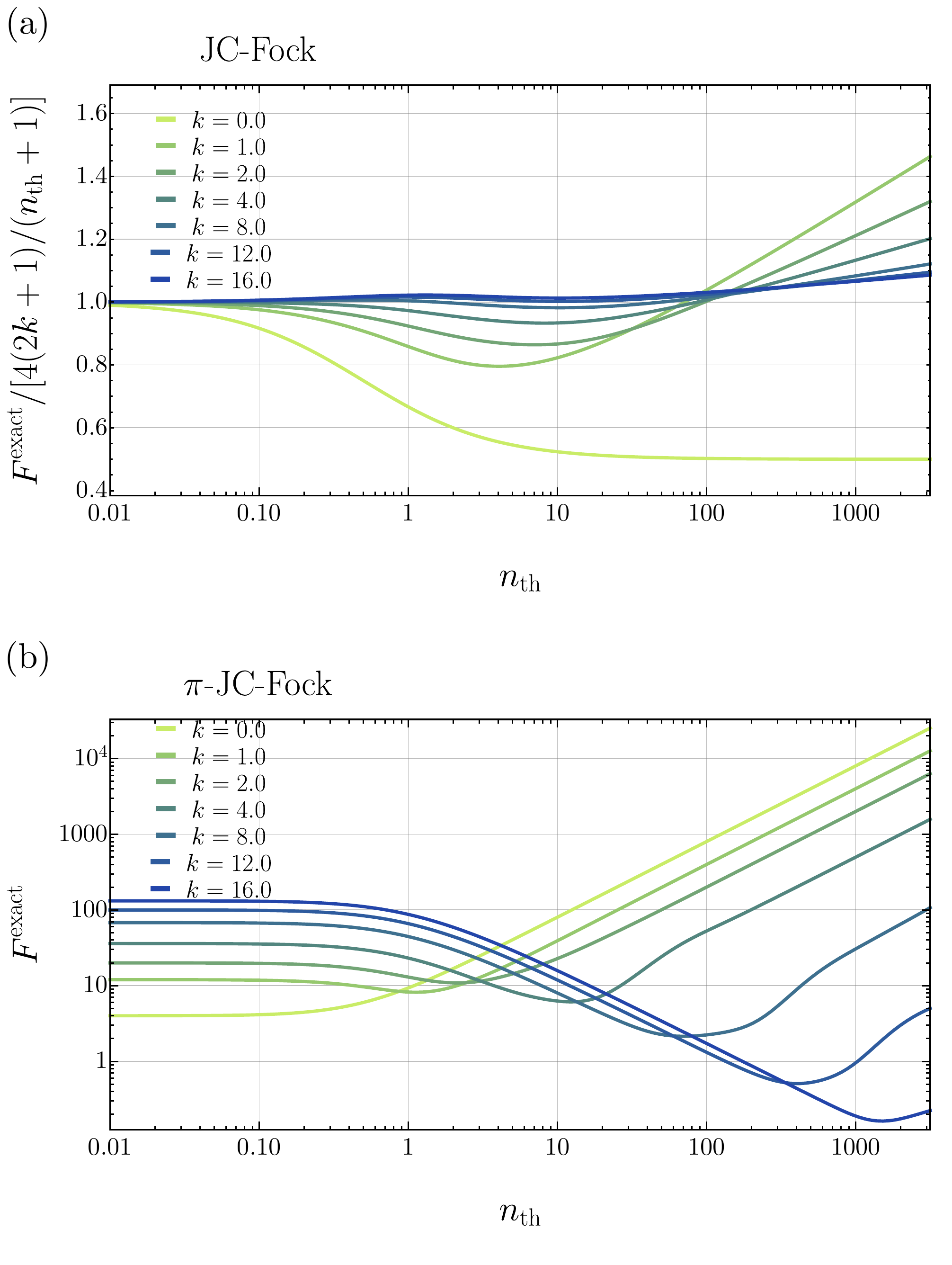}  
    \caption{Quantum Fisher information for displacement sensing using Jaynes--Cummings (JC) ladder-generated hot-Fock states. 
(a) Standard JC-generated hot-Fock states. 
The plotted quantity is normalized by the empirical asymptotic estimate used to describe the numerical data. 
The results remain close to the predicted scaling over a broad range of temperatures, confirming that calibrated JC ladder climbing does not remove the thermal mixed-state suppression. 
(b) Parity-filtered JC-generated hot-Fock states. 
For shallow ladders, parity filtering initially enhances the QFI by isolating a single parity ladder. 
However, calibrated JC pulses probabilistically transfer population between neighboring Fock states and therefore progressively repopulate the opposite parity sector. 
As a result, deep JC ladders eventually destroy the parity-protected enhancement and restore strong thermal suppression.
  \label{FigJCFock}}
\end{figure}

\smsubsubsection{Numerical approximation}
To further analyze the thermal suppression inherent to calibrated JC ladder climbing, we evaluate exact numerical expressions, which are shown in Fig.~\ref{FigJCFock}(a).
The numerical data are well described by the compact interpolation
\begin{align}
    \QFisher_{\JCLetter}^{(k)}
    \simeq
    \frac{
    4(2k+1)
    }{
    1+\ThermalOccupation
    }.
    \label{eq:qfi_jc_empirical}
\end{align}
This expression is exact in the ground-state limit, where the calibrated ladder prepares the pure Fock state $\ket{k}$, and gives the observed large-temperature scaling
\begin{align}
    \QFisher_{\JCLetter}^{(k)}
    \sim
    \frac{
    8k+4
    }{
    \ThermalOccupation
    }.
\end{align}
Note that we use Eq.~\eqref{eq:qfi_jc_empirical} only as an empirical asymptotic estimate, since the exact result is determined by the full formula~\eqref {eq:jc_path_sum}.

\smsubsubsection{Parity-filtered Jaynes--Cummings ladder climbing}
We now apply the calibrated JC ladder to an initially parity-filtered thermal state. 
Although the input occupies only one parity ladder, the calibrated JC sequence does not preserve this property. 
After the $j$th pulse, a component $\ket{n}$ transforms, after qubit reset, as in Eq.~\eqref{eq:nn_evolution},
\begin{align}
    \ket{n}\bra{n}
    \rightarrow
    \rounds{1-s_n^{(j)}}\ket{n}\bra{n}
    +
    s_n^{(j)}\ket{n+1}\bra{n+1}.
\end{align}
Each pulse creates population in both parity sectors unless $s_n^{(j)}$ is exactly zero or one. 
Consequently, after several pulses, the initially empty parity ladder is repopulated, and the parity-projected QFI identity no longer applies. 
Therefore, unlike parity-filtered phonon addition or deterministic number shifts, parity filtering does not protect the calibrated JC ladder: the ladder dynamics itself repopulates the opposite parity sector and restores mixed-state spectral suppression.

Numerically [see Fig.~\ref{FigJCFock}(b)], the parity-filtered JC ladder exhibits a competition between parity-enhanced displacement sensitivity and parity scrambling induced by imperfect ladder climbing. 
For the parity-filtered thermal state, the displacement operator initially connects the occupied parity ladder to an empty sector, leading to the enhanced scaling
\begin{align}
    \QFisher
    \sim
    8\ThermalOccupation .
\end{align}
However, each calibrated JC pulse probabilistically transfers
\begin{align}
    \ket{n}
    \rightarrow
    \ket{n}
    \quad \text{or} \quad
    \ket{n+1},
\end{align}
which necessarily mixes opposite parities.
The effect becomes stronger both for larger thermal occupation, which broadens the occupied Fock support, and for longer ladder sequences, which accumulate calibration mismatch over many steps.
Consequently, parity enhancement survives only for relatively shallow ladders, while deep calibrated JC ladders progressively restore mixed-state spectral suppression.
The exemplary state is shown in Fig.~\ref{FigSWigners}(f).

\smsubsection{SNAP-synthesized Susskind--Glogower shift}
The ideal Susskind--Glogower shift is
\begin{align}
    \ShiftOperator
    =
    \sum_{n=0}^{\infty}
    \ket{n+1}\bra{n}.
\end{align}
It obeys
\begin{align}
    \ShiftOperator^{\dagger}\ShiftOperator
    =
    \hat{\mathds{1}},
    \quad
    \ShiftOperator\ShiftOperator^{\dagger}
    =
    \hat{\mathds{1}}
    -
    \ket{0}\bra{0}.
\end{align}
Hence, it is not a unitary on the oscillator alone.
It can nevertheless be synthesized approximately in a finite subspace with SNAP-type control.
The ideal $k$-step action is
\begin{align}
    \ShiftOperator^k\ket{n}
    =
    \ket{n+k}.
\end{align}
Applied to the thermal state, it reads
\begin{align}
    \HotFockState_\SGLetter^{(k)}
    =
    \ShiftOperator^k
    \ThermalState
    \rounds{\ShiftOperator^{\dagger}}^k
    =
    \sum_{n=0}^{\infty}
    p_n
    \ket{n+k}\bra{n+k}.
\end{align}
Therefore,
\begin{align}
    p_{m,\SGLetter}^{(k)}
    =
    p_{m-k}
    \quad
    m\geq k,
\end{align}
and $p_{m,\SGLetter}^{(k)}=0$ for $m<k$.
The mean occupation is
\begin{align}
    \angles{\NumberOperator}_\SGLetter
    =
    \ThermalOccupation+k.
\end{align}
Thus, the mean occupation increases by exactly $k$, without the postselected Bose enhancement of Eq.~\eqref{eq:mean_phonon_added}.
Using Eq.~\eqref{eq:qfi_diagonal},
\begin{align}
    \QFisher_\SGLetter^{(k)}
    &=
    4k p_0
    +
    4
    \sum_{n=0}^{\infty}
    \rounds{n+k+1}
    \frac{
    \rounds{p_n-p_{n+1}}^2
    }{
    p_n+p_{n+1}
    }.
\end{align}
Since $p_0=1-\ResPar$ and $p_{n+1}=\ResPar p_n$,
\begin{align}
    \QFisher_\SGLetter^{(k)}
    &=
    4k\rounds{1-\ResPar}+
    4
    \frac{
    \rounds{1-\ResPar}^2
    }{
    1+\ResPar
    }
    \sum_{n=0}^{\infty}
    \rounds{n+k+1}p_n
    \nonumber\\
    &=
    4k\rounds{1-\ResPar}
    +
    4
    \frac{
    \rounds{1-\ResPar}^2
    }{
    1+\ResPar
    }
    \rounds{\ThermalOccupation+k+1}.
\end{align}
In terms of $\ThermalOccupation$,
\begin{align}
    \QFisher_\SGLetter^{(k)}
    =
    \frac{
    4k
    }{
    1+\ThermalOccupation
    }
    +
    \frac{
    4\rounds{\ThermalOccupation+k+1}
    }{
    \rounds{1+\ThermalOccupation}
    \rounds{1+2\ThermalOccupation}
    }.
    \label{eq:qfi_sg}
\end{align}
For $\ThermalOccupation=0$, this again gives $4\rounds{2k+1}$.
The exemplary state is shown in Fig.~\ref{FigSWigners}(g).

\smsubsubsection{Parity-filtered SNAP-synthesized Susskind--Glogower shift}
We now consider parity-filtered number shifting implemented through SNAP-synthesized Susskind--Glogower operators.
The SG operator changes the boson number by one,
\begin{align}
    \ParityOp
    \ShiftOperator
    \ParityOp
    =
    -
    \ShiftOperator,
\end{align}
hence
\begin{align}
    \ParityOp
    \ShiftOperator^k
    \ParityOp
    =
    \rounds{-1}^k
    \ShiftOperator^k .
\end{align}
Then, parity-filtered SG shifting again occupies only one parity sector:
\begin{itemize}
    \item for even $k$, the parity sector is preserved,
    \item for odd $k$, the parity sector is globally flipped.
\end{itemize}
Therefore,
\begin{align}
    \DensityMatrix_{k,\pm}^\SGLetter
    =
    \Projector_{\pm(-1)^k}
    \DensityMatrix_{k,\pm}^\SGLetter
    \Projector_{\pm(-1)^k}.
\end{align}
Again, displacement-sensing QFI simplifies exactly to
\begin{align}
    \QFisher_{k,\pm}^\SGLetter
    =
    4
    \Trace{
    \DensityMatrix_{k,\pm}^\SGLetter
    \rounds{
    \Creation+\Annihilation
    }^2
    },
\end{align}
which follows from the fact that the SG shift deterministically increases the occupation number by $k$,
\begin{align}
    \angles{
    \Creation\Annihilation
    }_{k,\pm}
    =
    k
    +
    \angles{
    \Creation\Annihilation
    }_{\pm},
\end{align}
and
\begin{align}
    \QFisher_{k,\pm}^\SGLetter
    =
    4
    \rounds{
    2k
    +
    2
    \angles{
    \Creation\Annihilation
    }_{\pm}
    +1
    }.
\end{align}
For the even-parity thermal state,
\begin{align}
    \angles{
    \Creation\Annihilation
    }_{+}
    =
    \frac{
    2\ThermalOccupation^2
    }{
    2\ThermalOccupation+1
    },
\end{align}
which gives
\begin{align}
    \QFisher_{k,+}^\SGLetter
    =
    4
    \rounds{
    2k
    +
    \frac{
    4\ThermalOccupation^2
    }{
    2\ThermalOccupation+1
    }
    +1
    }.
\end{align}
For large thermal occupation, $\ThermalOccupation\gg1$, the asymptotic scaling becomes
\begin{align}
    \QFisher_{k,+}^\SGLetter
    \simeq
    8\ThermalOccupation
    +
    8k.
\end{align}
Again, parity filtering qualitatively changes the role of SG ladder shifting.
Unlike Jaynes--Cummings ladder climbing, the ideal SG shift does not suffer from number-dependent Rabi frequencies and therefore avoids thermal dephasing between different Fock components.
In practice, however, approximate implementations based on SNAP gates and finite control sequences can still accumulate phase errors and truncation effects for broad thermal distributions.
The exemplary state is shown in Fig.~\ref{FigSWigners}(h).

\smsection{Hot cat states}

\smsubsection{ECD: superposition of displacements}
The echoed conditional displacement (ECD) protocol generates a cat state by first correlating the bosonic mode with an auxiliary two-level system and then erasing the auxiliary which-branch information.
In an ideal rotating-frame description, the elementary spin-dependent force can be written as
\begin{align}
    \Hamiltonian_\ECDLetter
    =
    \ImagUnit \hbar \JCCoup
    \PauliZ
    \rounds{
    \Creation-\Annihilation
    } .
\end{align}
For an interaction time $\DimTime$, this produces the conditional displacement unitary
\begin{align}
    \Unitary_{\ECDLetter}(\DimTime)
    =
    &\exp\rounds{
    \PauliZ
    \rounds{
    \CohDisp \Creation
    -
    \CohDisp \Annihilation
    }
    }
    = \nonumber \\
    &
    \ket{\uparrow}\bra{\uparrow}
    \Displacement{\CohDisp}
    +
    \ket{\downarrow}\bra{\downarrow}
    \Displacement{-\CohDisp},
\end{align}
where $\CohDisp=\JCCoup\DimTime$ is taken real. Starting from the auxiliary superposition
\begin{align}
    \ket{+}
    =
    \frac{
    \ket{\uparrow}+\ket{\downarrow}
    }{
    \sqrt{2}
    },
\end{align}
and an arbitrary bosonic state $\DensityMatrix$, the conditional displacement gives
\begin{align}
    \DensityMatrix_\text{cd} = \Unitary_{\ECDLetter}
    \rounds{
    \ket{+}\bra{+}
    \otimes
    \DensityMatrix
    }
    \Unitary_{\ECDLetter}^{\dagger}.
\end{align}
Projection of the auxiliary onto
\begin{align}
    \ket{+}_{\DispPhase}
    =
    \frac{
    \ket{\uparrow}
    +
    e^{-\ImagUnit\DispPhase}
    \ket{\downarrow}
    }{
    \sqrt{2}
    }
\end{align}
leaves the bosonic mode in the unnormalized state
\begin{align}
    \DensityMatrix_\ECDLetter \sim \ECDOperator
    \DensityMatrix_\text{cd}
    \ECDOperator^{\dagger},
\end{align}
with the Kraus operator
\begin{align}
    \ECDOperator
    =
    \Displacement{\CohDisp}
    +
    e^{\ImagUnit\DispPhase}
    \Displacement{-\CohDisp},
    \label{eq:ecd_operator_derivation}
\end{align}
up to a normalization factor.
The echo in the ECD sequence cancels unwanted unconditional rotations and geometric phases, leaving precisely this coherent sum of two opposite displacements in the ideal limit.
This is the key distinction from parity-filtering protocols: ECD does not remove one parity sector of the initial state, but coherently splits the entire initial state into two displaced branches.

For an initial thermal state, the normalized ECD hot cat is therefore
\begin{align}
    \HotCatState_\ECDLetter
    =
    \frac{
    \ECDOperator
    \ThermalState
    \ECDOperator^{\dagger}
    }{
    \Trac\squares{
    \ECDOperator
    \ThermalState
    \ECDOperator^{\dagger}
    }
    }.
    \label{eq:ecd_hot_state}
\end{align}
For arbitrary $\CohDisp$ and $\ThermalOccupation$ this state is mixed and non-Gaussian. Its exact QFI is obtained from Eq.~\eqref{eq:qfi_general} after diagonalization.
Let us now derive the large-separation approximation.
The large-separation regime is given by
\begin{align}
    \CohDisp^2
    \gg
    2\ThermalOccupation+1,
    \label{eq:ecd_large_separation}
\end{align}
where the two displaced thermal lobes are approximately orthogonal.
More explicitly, each Fock component of the thermal mixture is mapped to a superposition of two displaced Fock states,
\begin{align}
    \ket{n,\CohDisp}_{\DispPhase}
    =
    \frac{
    \Displacement{\CohDisp}\ket{n}
    +
    e^{\ImagUnit\DispPhase}
    \Displacement{-\CohDisp}\ket{n}
    }{
    \sqrt{2}
    },
\end{align}
where corrections to the normalization are exponentially small in the branch overlap
\begin{align}
    \matrixelement{n}{
    \Displacement{-2\CohDisp}
    }{
    n
    }
    =
    e^{-2\CohDisp^2}
    L_n\rounds{4\CohDisp^2}.
\end{align}
The ECD-cat is then approximately a thermal mixture over the internal index $n$, but each component carries the same coherent branch superposition,
\begin{align}
    \HotCatState_\ECDLetter
    \simeq
    \sum_n
    p_n
    \ket{n,\CohDisp}\bra{n,\CohDisp}_{\DispPhase}.
    \label{eq:ecd_mixture_branch_basis}
\end{align}
This structure is crucial: the mixedness remains in the internal thermal index, whereas the branch degree of freedom is pure and balanced for every occupied thermal component.
To make this separation explicit, let us introduce a local branch basis for each $n$,
\begin{align}
    \ket{R,n}
    =
    \Displacement{\CohDisp}\ket{n},
    \qquad
    \ket{L,n}
    =
    \Displacement{-\CohDisp}\ket{n}.
\end{align}
In the limit of~\eqref{eq:ecd_large_separation}, these states define an approximate tensor-product structure
\begin{align}
    \ket{R,n}
    \simeq
    \ket{R}_{\mathrm{b}}
    \otimes
    \ket{n}_{\mathrm{int}},
    \qquad
    \ket{L,n}
    \simeq
    \ket{L}_{\mathrm{b}}
    \otimes
    \ket{n}_{\mathrm{int}},
\end{align}
where the label $\mathrm{b}$ denotes the branch two-level system and $\mathrm{int}$ denotes fluctuations inside each lobe.
In this representation, the annihilation operator acts as
\begin{align}
    \Annihilation
    =
    \CohDisp
    \PauliZ^{\mathrm{b}}
    +
    \Annihilation_{\mathrm{int}}
    +
    \mathcal{O}\rounds{
    e^{-2\CohDisp^2}
    },
\end{align}
where $\PauliZ^{\mathrm{b}}\ket{R}_{\mathrm{b}}=\ket{R}_{\mathrm{b}}$ and $\PauliZ^{\mathrm{b}}\ket{L}_{\mathrm{b}}=-\ket{L}_{\mathrm{b}}$.
Consequently, for the displacement generator $\Quadrature$, we obtain
\begin{align}
    \Quadrature
    \simeq
    2\CohDisp\PauliZ^{\mathrm{b}}
    +
    \delta\Quadrature,
    \qquad
    \delta\Quadrature
    =
    \Annihilation_{\mathrm{int}}
    +
    \Creation_{\mathrm{int}}.
    \label{eq:ecd_generator_decomposition}
\end{align}
Eq.~\eqref{eq:ecd_generator_decomposition} is not an operator identity on the full Hilbert space.
Rather, it is the leading large-separation representation inside the subspace spanned by the two displaced thermal lobes.
The branch contribution is now obtained directly.
For each $n$, the ECD operation prepares a balanced branch superposition,
\begin{align}
    \ket{+}_{{\DispPhase},\mathrm{b}}
    =
    \frac{
    \ket{R}_{\mathrm{b}}
    +
    e^{\ImagUnit\DispPhase}
    \ket{L}_{\mathrm{b}}
    }{
    \sqrt{2}
    }.
\end{align}
For the phase choice giving maximal perpendicular-displacement sensitivity,
\begin{align}
    \angles{\PauliZ^{\mathrm{b}}}
    =
    0,
    \qquad
    \Variance\rounds{
    \PauliZ^{\mathrm{b}}
    }
    =
    1.
\end{align}
Therefore
\begin{align}
    4\Variance\rounds{
    2\CohDisp\PauliZ^{\mathrm{b}}
    }
    =
    16\CohDisp^2.
    \label{eq:ecd_branch_qfi}
\end{align}
This term is independent of $\ThermalOccupation$ because the same branch qubit is coherently prepared for every thermal component.
The internal contribution comes from $\delta\Quadrature$, which acts only inside each displaced thermal lobe.
Since the internal state is thermal, its displacement QFI is exactly
\begin{align}
    \QFisher_{\mathrm{int}}
    =
    \QFisher_\ThLetter
    =
    \frac{
    4
    }{
    2\ThermalOccupation+1
    }.
    \label{eq:ecd_internal_qfi}
\end{align}
The cross terms between the branch and internal generators vanish in the large-separation approximation because the branch state has no bias, $\angles{\PauliZ^{\mathrm{b}}}=0$,
and the thermal internal state has no first moments, $ \angles{\delta\Quadrature}_{\ThermalState}=0$.
Combining Eqs.~\eqref{eq:ecd_branch_qfi} and~\eqref{eq:ecd_internal_qfi}, we obtain
\begin{align}
    \QFisher_{\text{ECD}}
    \simeq
    16\CohDisp^2
    +
    \frac{
    4
    }{
    2\ThermalOccupation+1
    }.
    \label{eq:qfi_ecd}
\end{align}
This is an asymptotic large-separation result.
It shows that the ECD hot cat preserves a pure branch-interference contribution even though the initial state is thermal.
The thermal mixedness remains confined to the internal lobe degrees of freedom and only contributes to the local thermal term.
This is why ECD differs qualitatively from parity-engineered protocols: its leading sensitivity comes from coherent interference between displaced thermal branches rather than from projection onto a single parity sector.
The exemplary state is shown in Fig.~\ref{FigSWigners}(i).

\begin{figure}[ht!]
    \includegraphics[width=\linewidth]{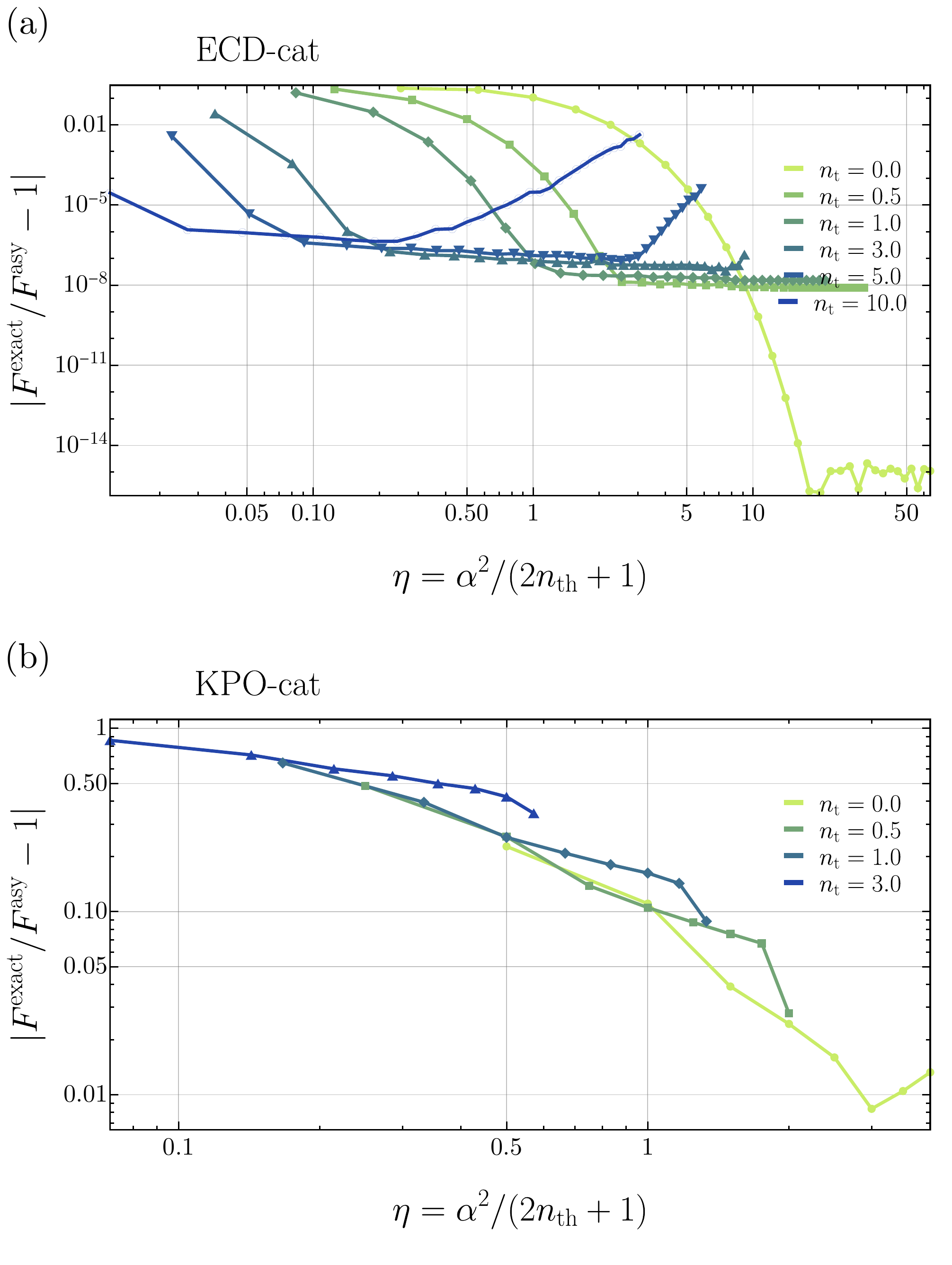}  
    \caption{Numerical validation of the asymptotic large-separation formulas for hot cat states. 
(a) Relative error between the exact mixed-state QFI and the asymptotic ECD expression in Eq.~\eqref{eq:qfi_ecd} as a function of the dimensionless separation parameter $\eta$.
The exact QFI is computed from the full spectral formula in Eq.~\eqref{eq:qfi_general}, without assuming orthogonality of the displaced thermal branches. The systematic collapse of the curves confirms that the asymptotic accuracy is controlled predominantly by the branch-separation parameter $\eta$. The remaining deviations originate from finite overlap corrections between the two displaced thermal lobes. 
(b) Analogous comparison for KPO-generated hot cat states obtained from full parity-preserving KPO dynamics under the Hamiltonian in Eq.~\eqref{eq:kpo_hamiltonian}. The asymptotic KPO expression becomes increasingly accurate with increasing $\eta$, confirming the emergence of the large-separation doublet structure associated with the two semiclassical KPO wells. Residual deviations arise from finite branch overlap and finite-time nonadiabatic corrections during the KPO ramp.
  \label{FigS1}}
\end{figure}

\begin{figure*}[ht!]
    \includegraphics[width=\linewidth]{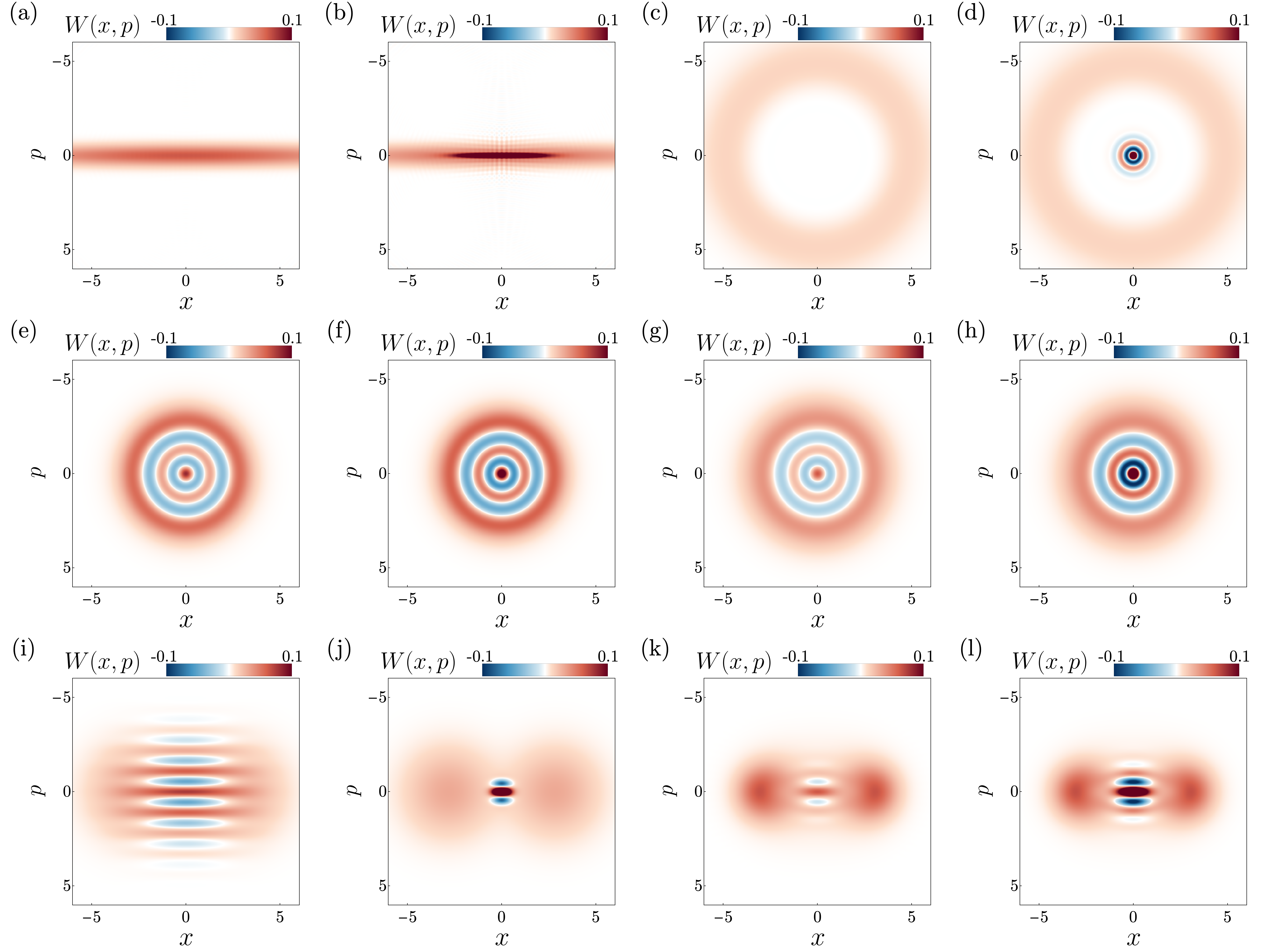}  
    \caption{Wigner functions of the hot states considered in this work for $\ThermalOccupation=2$ and 4 added excitations.
    Panels (a,b) show squeezed thermal states without and with parity filtering. 
    Panels (c,d) show phonon-added thermal states without and with parity filtering. 
    Panels (e,f) show Jaynes--Cummings ladder-climbed states without and with parity filtering. 
    Panels (g,h) show SNAP-synthesized Susskind--Glogower shifted states without and with parity filtering. 
    Panels (i)--(l) show hot cat states: (i) ECD, (j) qcMAP, (k) KPO catification, and (l) parity-filtered KPO catification. 
  \label{FigSWigners}}
\end{figure*}

\smsubsubsection{Numerical check}
To verify the validity of the asymptotic ECD expression, we numerically compare the exact mixed-state QFI with the analytical large-separation formula derived from the displaced-branch approximation.
The ECD-cat is generated by coherent superposition of opposite displacements applied directly to the thermal state.
The exact QFI is evaluated numerically from the full mixed-state spectral expression in Eq.~\eqref{eq:qfi_general}, without assuming orthogonality between the displaced thermal branches.
We compare it with the asymptotic prediction in Eq.~\eqref{eq:qfi_ecd}. The relevant control parameter is
\begin{align}
    \eta
    =
    \frac{
    \CohDisp^2
    }{
    2\ThermalOccupation+1
    },
\end{align}
which measures the separation of the two displaced thermal lobes relative to their thermal width.
Physically, the branch overlap is exponentially suppressed as $\sim \exp\rounds{-2\eta}$.
Figure~\ref{FigS1}(a) shows the relative error between the exact and asymptotic QFI as a function of $\eta$ for different thermal occupations.
The approximate collapse of the curves confirms that the displaced-branch approximation is controlled predominantly by the dimensionless separation parameter $\eta$, while the remaining deviations originate from finite overlap corrections between the two thermal branches.

\smsubsection{qcMAP: parity filtering after displacement}
Let us start with the ideal qubit-cavity mapping (qcMAP) operation with outcome $\ParityOutcome=\pm1$,
\begin{align}
    \QCMapOperator^{\ParityOutcome}
    =
    \ParityProjector
    \Displacement{\CohDisp}.
\end{align}
After the action on the thermal state, the normalized state is then 
\begin{align}
    \HotCatState_{\QCLetter}^{\ParityOutcome}
    =
    \frac{
    \ParityProjector
    \Displacement{\CohDisp}
    \ThermalState
    \DisplacementDagger{\CohDisp}
    \ParityProjector
    }{
    p_{\ParityOutcome}
    },
    \label{eq:qcmap_state}
\end{align}
where the outcome probability is
\begin{align}
    p_{\ParityOutcome} = \Tr[ \ParityProjector
    \Displacement{\CohDisp}
    \ThermalState
    \DisplacementDagger{\CohDisp}
    \ParityProjector
       ].
\end{align}
The form of Eq.~\eqref{eq:qcmap_state} should be understood as an effective description of the idealized qcMAP protocol in the large-separation regime.
In the experimentally implemented protocol, the cavity mode interacts dispersively with an ancilla qubit through
\begin{align}
    \Hamiltonian_\DispStrength
    =
    \DispStrength
    \Creation \Annihilation
    \PauliZ,
\end{align}
followed by conditional cavity displacements and selective ancilla rotations.
Here, $\DispStrength$ is the coupling strength and $\PauliZ$ is the Pauli $z$ operator on the qubit space.
The protocol does not perform an explicit projective measurement of the cavity parity.
Instead, the dispersive evolution maps different parity sectors onto different ancilla-conditioned phase-space trajectories.
In the large-cat regime, this sequence approximately realizes a parity-conditioned displacement of the form
\begin{align}
    \Displacement{\CohDisp}\Projector_{+}
    +
    \Displacement{-\CohDisp}\Projector_{-},
\end{align}
which spatially separates the even- and odd-parity sectors in phase space.
Postselection on the ancilla outcome therefore effectively isolates a single parity sector, leading to the parity-filtered structure in Eq.~\eqref{eq:qcmap_state}.
As a consequence, the metrological enhancement of qcMAP-cats originates from approximate parity-sector engineering rather than from coherent interference between displaced thermal branches.
Using
\begin{align}
    \Trac\squares{
    \ParityOp
    \Displacement{\CohDisp}
    \ThermalState
    \DisplacementDagger{\CohDisp}
    }
    =
    \frac{
    1
    }{
    2\ThermalOccupation+1
    }
    \exp\rounds{
    -
    \frac{
    2\CohDisp^2
    }{
    2\ThermalOccupation+1
    }
    },
\end{align}
we obtain
\begin{align}
    p_{\ParityOutcome}
    =
    \frac{
    1+\ParityOutcome \chi_{\Parity}
    }{
    2
    },
    \quad
    \chi_{\Parity}
    =
    \frac{
    1
    }{
    2\ThermalOccupation+1
    }
    \exp\rounds{
    -
    \frac{
    2\CohDisp^2
    }{
    2\ThermalOccupation+1
    }
    }.
\end{align}
The state in Eq.~\eqref{eq:qcmap_state} is mixed, but it has support only in one parity sector.
Again, for parity-filtered states we have
\begin{align}
    \QFisher_\QCLetter^{\ParityOutcome}
    =
    4
    \Tr(
    \HotCatState_\QCLetter^{\ParityOutcome}
    \Quadrature^2
    ).
    \label{eq:qfi_parity_support}
\end{align}
We now compute the trace. First,
\begin{align}
    \Trac\squares{
    \Displacement{\CohDisp}
    \ThermalState
    \DisplacementDagger{\CohDisp}
    \Quadrature^2
    }
    =
    4\CohDisp^2
    +
    2\ThermalOccupation+1.
\end{align}
Second,
\begin{align}
    \Trac\squares{
    \Displacement{\CohDisp}
    \ThermalState
    \DisplacementDagger{\CohDisp}
    \ParityOp \Quadrature^2
    }
    =
    \frac{
    \chi_{\Parity}
    }{
    2\ThermalOccupation+1
    }.
\end{align}
Combining these with Eq.~\eqref{eq:qfi_parity_support} yields the ideal qcMAP result
\begin{align}
    \QFisher_\QCLetter^{\ParityOutcome}
    =
    4
    \frac{
    4\CohDisp^2
    +
    2\ThermalOccupation+1
    +
    \ParityOutcome
    \chi_{\Parity}/\rounds{2\ThermalOccupation+1}
    }{
    1+\ParityOutcome\chi_{\Parity}
    }.
    \label{eq:qfi_qcmap_exact}
\end{align}
In the large-separation regime $\chi_{\Parity}\rightarrow0$,
\begin{align}
    \QFisher_\QCLetter
    \simeq
    16\CohDisp^2
    +
    4\rounds{2\ThermalOccupation+1}.
    \label{eq:qfi_qcmap_large}
\end{align}
The growth of the second term is a consequence of parity filtering.
The displacement generator moves the state into an initially empty parity sector.
Equation~\eqref{eq:qfi_qcmap_large} can also be understood from the general structure of states with definite parity.
Since the qcMAP state occupies only one parity sector and the displacement generator changes parity, we have Eq.~\eqref{eq:qfi_parity_only},
\begin{align}
    \QFisher_\QCLetter
    =
    4
    \rounds{
    2\angles{\Creation \Annihilation}
    +
    1
    +
    \angles{\Annihilation^2+{\Annihilation}^{\dagger 2}}
    }.
    \label{eq:qfi_parity_general}
\end{align}
For large cat separation, interference between the two branches becomes exponentially suppressed inside local observables, and
\begin{align}
    \angles{\Annihilation^2+{\Annihilation}^{\dagger 2}}
    \simeq
    2\CohDisp^2.
\end{align}
At the same time, the mean occupation becomes
\begin{align}
    \angles{\Creation \Annihilation}
    \simeq
    \ThermalOccupation
    +
    \CohDisp^2.
\end{align}
Substituting into Eq.~\eqref{eq:qfi_parity_general} gives
\begin{align}
    \QFisher_{\text{qcMAP}}
    &\simeq
    4
    \rounds{
    2\rounds{
    \ThermalOccupation+\CohDisp^2
    }
    +
    1
    +
    2\CohDisp^2
    }
    \nonumber\\
    &=
    16\CohDisp^2
    +
    4\rounds{
    2\ThermalOccupation+1
    }.
\end{align}
Thus, the thermal enhancement is fully consistent with the generic structure of definite-parity states.
The additional factor proportional to $\CohDisp^2$ originates from the anomalous second moments, $\angles{\Annihilation^2}$ and $\angles{{\Annihilation}^{\dagger 2}}$, generated by the cat-state coherence.
Additionally, the success probability of the qcMAP preparation in the large-separation limit is
\begin{align}
    p_{+}
    \simeq
    p_{-}
    \simeq
    \frac{1}{2}.
\end{align}
Thus, the protocol does not suffer from exponentially small preparation probability, which is often typical for heralded protocols.
Since a relevant figure for conditional metrology is the preparation-probability-weighted QFI, in the large-separation limit we have
\begin{align}
    p_{\ParityOutcome}
    \QFisher_\QCLetter^{\ParityOutcome}
    \simeq
    8\CohDisp^2
    +
    2\rounds{2\ThermalOccupation+1}.
\end{align}
Therefore, the thermal enhancement survives even after accounting for the probabilistic nature of the parity filtering.
The exemplary state is shown in Fig.~\ref{FigSWigners}(j).

\smsubsection{KPO catification}

We now consider the adiabatic generation of cat states using a Kerr parametric oscillator (KPO).
The KPO Hamiltonian in the rotating frame is
\begin{align}
    \Hamiltonian_\KPOLetter
    =
    \frac{\Kerr}{2}
    \Annihilation^{\dagger 2}
    \Annihilation^2
    -
    \frac{\PumpAmp}{2}
    \rounds{
    \Annihilation^{\dagger 2}
    +
    \Annihilation^2
    },
    \label{eq:kpo_hamiltonian}
\end{align}
with Kerr nonlinearity $\Kerr$ and two-photon pump amplitude $\PumpAmp$.
The Hamiltonian preserves parity,
\begin{align}
    \commutator{
    \Hamiltonian_\KPOLetter
    }{
    \ParityOp
    }
    =
    0,
\end{align}
and therefore the adiabatic evolution decomposes into independent even and odd sectors.
The adiabatic mapping between vacuum/one-photon states and even/odd cat states is well established in the Kerr-cat literature~\cite{Puri2020,Frattini2024}.
In addition, the excited-state spectrum of the KPO forms pairwise doublets associated with local excitations around the two semiclassical wells~\cite{Frattini2024,Frattini2021}.
Motivated by this double-well structure, we derive below the asymptotic large-pump mapping for arbitrary Fock states.
For
\begin{align}
    \CohDisp^2
    =
    \frac{
    \PumpAmp
    }{
    \Kerr
    },
\end{align}
the Hamiltonian can be rewritten as
\begin{align}
    \Hamiltonian_{\KPOLetter}
    =
    \frac{\Kerr}{2}
    \rounds{
    \Annihilation^{\dagger 2}-\CohDisp^2
    }
    \rounds{
    \Annihilation^2-\CohDisp^2
    }
    -
    \frac{\Kerr}{2}\CohDisp^4.
\end{align}
The classical minima therefore satisfy
\begin{align}
    \Annihilation
    =
    \pm
    \CohDisp.
\end{align}
We now expand around the right minimum,
\begin{align}
    \Annihilation
    =
    \CohDisp+\hat{b}.
\end{align}
Then
\begin{align}
    \Annihilation^2-\CohDisp^2
    =
    2\CohDisp \hat{b}
    +
    \hat{b}^2.
\end{align}
Keeping only quadratic terms in the fluctuations gives
\begin{align}
    \Hamiltonian_{\KPOLetter}
    \simeq
    2\Kerr \CohDisp^2
    \hat{b}^{\dagger}\hat{b}
    -
    \frac{\Kerr}{2}\CohDisp^4.
\end{align}
Thus the local excitations around each well are approximately harmonic oscillator states.
The corresponding local eigenstates are $\Displacement{\CohDisp}\ket{m}$ and $\Displacement{-\CohDisp}\ket{m}$.
The overlap between the two wells is exponentially suppressed,
\begin{align}
    \bra{m}
    \Displacement{-2\CohDisp}
    \ket{m}
    =
    e^{-2\CohDisp^2}
    L_m\rounds{4\CohDisp^2},
\end{align}
and therefore each local excitation level forms an approximate doublet.
The exact eigenstates must additionally have definite parity. 
Using
\begin{align}
    \ParityOp
    \Displacement{\CohDisp}
    \ParityOp
    =
    \Displacement{-\CohDisp},
\end{align}
together with
\begin{align}
    \ParityOp
    \ket{m}
    =
    \rounds{-1}^m
    \ket{m},
\end{align}
we obtain
\begin{align}
    \ParityOp
    \Displacement{\CohDisp}
    \ket{m}
    =
    \rounds{-1}^m
    \Displacement{-\CohDisp}
    \ket{m}.
\end{align}
Consider now the superposition
\begin{align}
    \ket{\mathcal{C}_{m,s}}
    =
    \frac{
    \Displacement{\CohDisp}\ket{m}
    +
    s
    \rounds{-1}^m
    \Displacement{-\CohDisp}\ket{m}
    }{
    \sqrt{2}
    },
    \qquad
    s=\pm1.
    \label{eq:excited_cat_correct}
\end{align}
Applying parity gives
\begin{align}
    \ParityOp
    \ket{\mathcal{C}_{m,s}}
    =
    s
    \ket{\mathcal{C}_{m,s}}.
\end{align}
Therefore $\ket{\mathcal{C}_{m,+}}$ has even parity, while $\ket{\mathcal{C}_{m,-}}$ has odd parity independently of $m$.
Since parity is conserved during the adiabatic ramp and levels within the same parity sector do not cross, adiabatic continuity implies the asymptotic mapping
\begin{align}
    \ket{2m}
    \rightarrow
    \ket{\mathcal{C}_{m,+}},
    \qquad
    \ket{2m+1}
    \rightarrow
    \ket{\mathcal{C}_{m,-}}.
    \label{eq:kpo_mapping}
\end{align}
Consequently, the effective large-cat unitary becomes
\begin{align}
    \KPOUnitary
    \simeq
    \sum_{m=0}^{\infty}
    \ket{\mathcal{C}_{m,+}}
    \bra{2m}
    +
    \sum_{m=0}^{\infty}
    \ket{\mathcal{C}_{m,-}}
    \bra{2m+1}.
    \label{eq:kpo_unitary_correct}
\end{align}
Eq.~\eqref{eq:kpo_unitary_correct} should be understood as an asymptotic semiclassical large-pump description motivated by the double-well structure of the KPO Hamiltonian rather than an exact finite-$\CohDisp$ identity.
For the lowest states, one recovers the standard even/odd cat mapping,
\begin{align}
    \ket{0}
    &\rightarrow
    \frac{
    \ket{\CohDisp}
    +
    \ket{-\CohDisp}
    }{
    \sqrt{2}
    },
    \\
    \ket{1}
    &\rightarrow
    \frac{
    \ket{\CohDisp}
    -
    \ket{-\CohDisp}
    }{
    \sqrt{2}
    }.
\end{align}
However, the next states become
\begin{align}
    \ket{2}
    &\rightarrow
    \frac{
    \Displacement{\CohDisp}\ket{1}
    -
    \Displacement{-\CohDisp}\ket{1}
    }{
    \sqrt{2}
    },
    \\
    \ket{3}
    &\rightarrow
    \frac{
    \Displacement{\CohDisp}\ket{1}
    +
    \Displacement{-\CohDisp}\ket{1}
    }{
    \sqrt{2}
    }.
\end{align}
Thus higher Fock states map onto superpositions of displaced excited local oscillator states rather than ordinary cat states.
Applying Eq.~\eqref{eq:kpo_unitary_correct} to the thermal state gives
\begin{align}
    \HotCatState_{\KPOLetter}
    \simeq
    \sum_m
    p_{2m}
    \ket{\mathcal{C}_{m,+}}
    \bra{\mathcal{C}_{m,+}}
    +
    \sum_m
    p_{2m+1}
    \ket{\mathcal{C}_{m,-}}
    \bra{\mathcal{C}_{m,-}}.
    \label{eq:kpo_hot_state_correct}
\end{align}
The resulting state is therefore a mixture of even and odd excited-cat manifolds.
In the large-separation regime, similarly to the ECD case, the generator can again be decomposed into branch and local contributions,
\begin{align}
    \Quadrature
    \simeq
    2\CohDisp
    \PauliZ^{\text{b}}
    +
    \delta \Quadrature.
\end{align}
In contrast to qcMAP, the KPO evolution preserves the initial thermal parity weights.
The resulting state, therefore, remains an incoherent mixture of even- and odd-parity cat manifolds,
\begin{align}
    \HotCatState_{\KPOLetter}
    \simeq
    p_{\text{e}}
    \DensityMatrix_{\text{e}}
    +
    p_{\text{o}}
    \DensityMatrix_{\text{o}},
\end{align}
where
\begin{align}
    p_{\text{e}}
    =
    \sum_{m=0}^{\infty}
    p_{2m},
    \qquad
    p_{\text{o}}
    =
    \sum_{m=0}^{\infty}
    p_{2m+1}.
\end{align}
Using the thermal distribution,
\begin{align}
    p_n
    =
    \rounds{
    1-\ResPar
    }
    \ResPar^n,
\end{align}
we obtain
\begin{align}
    p_{\text{e}}
    &=
    \rounds{
    1-\ResPar
    }
    \sum_{m=0}^{\infty}
    \ResPar^{2m}
    =
    \frac{
    1
    }{
    1+\ResPar
    },
    \\
    p_{\text{o}}
    &=
    \rounds{
    1-\ResPar
    }
    \sum_{m=0}^{\infty}
    \ResPar^{2m+1}
    =
    \frac{
    \ResPar
    }{
    1+\ResPar
    }.
\end{align}
The thermal parity expectation value is therefore
\begin{align}
    \angles{\ParityOp}_\ThLetter
    &=
    p_{\text{e}}
    -
    p_{\text{o}} =
    \frac{
    1-\ResPar
    }{
    1+\ResPar
    }
    =
    \frac{
    1
    }{
    2\ThermalOccupation+1
    }.
    \label{eq:thermal_parity}
\end{align}
The branch part of the displacement generator does not have diagonal matrix elements within a definite-parity cat doublet.
Instead, it couples the two parity partners associated with the same local excitation index,
\begin{align}
    \matrixelement{\mathcal{C}_{m,+}}{\Quadrature}{\mathcal{C}_{m,-}}
    \simeq
    2\CohDisp,
\end{align}
up to exponentially small corrections in the branch overlap.
The branch contribution to the mixed-state QFI therefore comes from spectral pairs with weights $p_{2m}$ and $p_{2m+1}$:
\begin{align}
    \QFisher_{\KPOLetter,\mathrm{branch}}
    &\simeq
    16\CohDisp^2
    \sum_{m=0}^{\infty}
    \frac{
    \rounds{p_{2m}-p_{2m+1}}^2
    }{
    p_{2m}+p_{2m+1}
    }.
\end{align}
Using the thermal distribution $p_n=(1-\ResPar)\ResPar^n$, the sum evaluates to
\begin{align}
    \sum_{m=0}^{\infty}
    \frac{
    \rounds{p_{2m}-p_{2m+1}}^2
    }{
    p_{2m}+p_{2m+1}
    }
    =
    \rounds{
    \frac{1-\ResPar}{1+\ResPar}
    }^2
    =
    \angles{\ParityOp}_\ThLetter^2.
\end{align}
Hence
\begin{align}
    \QFisher_{\KPOLetter, \text{branch}}
    &\simeq
    16\CohDisp^2
    \angles{\ParityOp}_\ThLetter^2 = 
    \frac{
    16\CohDisp^2
    }{
    \rounds{
    2\ThermalOccupation+1
    }^2
    }.
\end{align}
Therefore, unlike qcMAP, the KPO protocol does not remove the incoherent mixing between even and odd cat sectors.
Increasing thermal occupation progressively equalizes the populations of the two parity ladders,
\begin{align}
    p_{\text{e}}
    \rightarrow
    p_{\text{o}}
    \rightarrow
    \frac12,
\end{align}
which suppresses the doublet spectral contrast and consequently reduces the cat-enhanced contribution to the QFI.
The local excitation index is now $m$. The induced distribution over local excitations becomes
\begin{align}
    P_m
    =
    p_{2m}+p_{2m+1}
    =
    \rounds{
    1-\ResPar
    }
    \rounds{
    1+\ResPar
    }
    \ResPar^{2m}.
\end{align}
This corresponds to an effective thermal distribution with parameter $\ResPar^2$. Using Eq.~\eqref{eq:qfi_thermal},
\begin{align}
    \QFisher_{\text{KPO,int}}
    =
    4
    \frac{
    1-\ResPar^2
    }{
    1+\ResPar^2
    } = 4\frac{1 + 2 \ThermalOccupation}{1 + 2 \ThermalOccupation (1 + \ThermalOccupation)}.
\end{align}
The resulting asymptotic QFI becomes
\begin{align}
    \QFisher_{\KPOLetter}
    \simeq
    \frac{
    16\CohDisp^2
    }{
    \rounds{
    2\ThermalOccupation+1
    }^2
    }
    +
    4
    \frac{
    1-\ResPar^2
    }{
    1+\ResPar^2
    }.
    \label{eq:qfi_kpo_correct}
\end{align}
The exemplary state is shown in Fig.~\ref{FigSWigners}(k).

\smsubsubsection{Parity selection followed by KPO catification}
We now consider first selecting a definite parity sector of the initial thermal state,
\begin{align}
    \ThermalState^\ParityOutcome
    =
    \frac{
    \ParityProjector
    \ThermalState
    \ParityProjector
    }{
    \Tr\rounds{
    \ParityProjector\ThermalState
    }
    }.
\end{align}
where $\ParityOutcome=\pm1$ labels the even and odd sectors.
For the even outcome,
\begin{align}
    \ThermalState^{+}
    =
    \rounds{
    1-\ResPar^2
    }
    \sum_{m=0}^{\infty}
    \ResPar^{2m}
    \ket{2m}\bra{2m},
\end{align}
while for the odd outcome,
\begin{align}
    \ThermalState^{-}
    =
    \rounds{
    1-\ResPar^2
    }
    \sum_{m=0}^{\infty}
    \ResPar^{2m}
    \ket{2m+1}\bra{2m+1}.
\end{align}
Here
\begin{align}
    \ResPar
    =
    \frac{
    \ThermalOccupation
    }{
    1+\ThermalOccupation
    }.
\end{align}
The KPO evolution preserves parity and adiabatically maps Fock states into excited cat eigenstates of the same symmetry.
Consequently, parity selection removes the incoherent mixture of even and odd cat sectors present for ordinary hot KPO cats.
After catification,
\begin{align}
    \HotCatState_{\ParityOutcome,\KPOLetter}
    =
    \rounds{
    1-\ResPar^2
    }
    \sum_{m=0}^{\infty}
    \ResPar^{2m}
    \ket{\mathcal{C}_{m,\ParityOutcome}}
    \bra{\mathcal{C}_{m,\ParityOutcome}},
    \label{eq:parity_kpo_state_rewritten2}
\end{align}
where all cat branches now share the same parity symmetry.
For states with definite parity, we have $\QFisher
    =
    4
    \rounds{
    2\angles{\Creation \Annihilation}
    +
    1
    +
    \angles{\Annihilation^2}
    +
    \angles{\Annihilation^{\dagger 2}}
    }$.
In the large-cat regime, each excited KPO-cat eigenstate contributes approximately
\begin{align}
    \angles{\Annihilation^2}
    \simeq
    \CohDisp^2,
    \qquad
    \angles{\Annihilation^{\dagger 2}}
    \simeq
    \CohDisp^2,
\end{align}
while the occupation decomposes into the branch and the thermal contributions,
\begin{align}
    \angles{\Creation \Annihilation}
    \simeq
    \CohDisp^2
    +
    \angles{\Creation \Annihilation}_{\ParityOutcome}.
\end{align}
Consequently,
\begin{align}
    \QFisher_{\ParityOutcome,\KPOLetter}
    \simeq
    4
    \rounds{
    4\CohDisp^2
    +
    2\angles{\Creation \Annihilation}_{\ParityOutcome}
    +
    1
    }.
    \label{eq:qfi_pf_kpo_general}
\end{align}
For the even sector,
\begin{align}
    \angles{\Creation \Annihilation}_{+}
    =
    \frac{
    2\ThermalOccupation^2
    }{
    2\ThermalOccupation+1
    },
\end{align}
which gives
\begin{align}
    \QFisher_{+,\KPOLetter}
    \simeq
    4
    \rounds{
    4\CohDisp^2
    +
    \frac{
    4\ThermalOccupation^2
    }{
    2\ThermalOccupation+1
    }
    +
    1
    }.
    \label{eq:qfi_pf_kpo_even_final}
\end{align}
For the odd sector,
\begin{align}
    \angles{\Creation \Annihilation}_{-}
    =
    \frac{
    2\ThermalOccupation^2
    +
    2\ThermalOccupation
    +
    1
    }{
    2\ThermalOccupation+1
    },
\end{align}
leading to
\begin{align}
    \QFisher_{-,\KPOLetter}
    \simeq
    4
    \rounds{
    4\CohDisp^2
    +
    2
    \frac{
    2\ThermalOccupation^2
    +
    2\ThermalOccupation
    +
    1
    }{
    2\ThermalOccupation+1
    }
    +
    1
    }.
    \label{eq:qfi_pf_kpo_odd_final}
\end{align}
Thus, parity filtering restores the coherent cat contribution suppressed in ordinary hot KPO cats, while simultaneously modifying the thermal contribution through the parity-projected occupation statistics.
This protocol is conceptually close to qcMAP since both recover coherent cat branches through parity selection.
However, the order and mechanism differ.
qcMAP first creates displaced thermal branches and subsequently projects onto a parity sector.
In contrast, parity selection followed by KPO catification first filters the thermal ensemble and then adiabatically maps each Fock component into an excited KPO-cat eigenstate.
As a result, qcMAP produces parity-filtered displaced thermal cats, whereas the present protocol generates parity-filtered mixtures of excited KPO-cat states.
The exemplary state is shown in Fig.~\ref{FigSWigners}(l).

\smsubsubsection{Numerical check}
To check the validity of the asymptotic KPO expression, we simulate the full parity-preserving KPO dynamics rather than imposing the large-cat mapping by hand.
The Hamiltonian in Eq.~\eqref{eq:kpo_hamiltonian} is ramped from $\PumpAmp=0$ to a final value $\PumpAmp_{\mathrm f}$, and the associated asymptotic cat amplitude is identified as
\begin{align}
    \CohDisp_{\mathrm{eff}}^2
    =
    \frac{
    \PumpAmp_{\mathrm f}
    }{
    \Kerr
    } .
\end{align}
Since the initial thermal state is diagonal in the Fock basis, we evolve each initially occupied number state independently under the unitary KPO ramp and reconstruct the final mixed state as
\begin{align}
    \DensityMatrix_{\KPOLetter}
    &=
    \sum_n
    \Prob_n
    \ket{\psi_n(T)}\bra{\psi_n(T)}, \nonumber \\
    \ket{\psi_n(T)}
    &=
    \Unitary_{\KPOLetter}(T)\ket{n}.
\end{align}
The exact QFI is then computed from the full mixed-state spectral formula in Eq.~\eqref{eq:qfi_general}, without assuming a two-branch approximation.
We compare this numerical value with the asymptotic prediction in Eq.~\eqref{eq:qfi_kpo_correct}.
The relevant control parameter is
\begin{align}
    \eta
    =
    \frac{
    \CohDisp_{\mathrm{eff}}^2
    }{
    2\ThermalOccupation+1
    },
\end{align}
which measures the separation of the two KPO wells relative to the thermal width of each lobe. Figure~\ref{FigS1}(b) shows the relative error between the exact and asymptotic QFI as a function of $\eta$.
The systematic decrease of the error with increasing $\eta$ confirms that the analytical expression is controlled by the large-separation limit, while the remaining deviations arise from finite branch overlap and finite-time nonadiabatic corrections during the KPO ramp.

\smsection{Instantaneous noise susceptibility of hot-state metrological resources}
We now analyze the susceptibility of the displacement QFI to noise during the sensing/storage stage. 
The state is assumed to be prepared ideally at $\DimTime=0$ and then evolves for a short time under a noisy bosonic channel before the displacement readout. 
The quantity of interest is the instantaneous QFI susceptibility
\begin{align}
    \DotQFisher
    =
    \left.
    \frac{
    \partial
    }{
    \partial \DimTime
    }
    \QFisher[
    \DensityMatrix(\DimTime),
    \Quadrature
    ]
    \right|_{\DimTime=0},
    \label{eq:instantaneous_qfi_susceptibility}
\end{align}
where $\DensityMatrix(\DimTime)$ is the probe state evolving under the noisy channel.
This quantity measures the initial degradation rate of the metrological resource, independently of the later-time nonlinear decay.

\smsubsection{Noise model}
We consider three elementary noise mechanisms:
\begin{align}
    \frac{\dd\DensityMatrix}{\dd\DimTime}
    =
    \DimBosonLossRate
    \Dissipator{\Annihilation}\DensityMatrix
    +
    \DimHeatingRate
    \rounds{
    \Dissipator{\Annihilation}
    +
    \Dissipator{\Creation}
    }
    \DensityMatrix
    +
    \DimBosonDephasingRate
    \Dissipator{\Creation\Annihilation}\DensityMatrix .
    \label{eq:noise_master_equation}
\end{align}
The first term describes zero-temperature bosonic loss.
The second describes phase-insensitive motional heating in the rotating-wave approximation.
Introducing the conjugate quadrature $\QuadratureP=\ImagUnit(\Creation-\Annihilation)$, it can be written as
\begin{align}
    \Dissipator{\Annihilation}
    +
    \Dissipator{\Creation}
    =
    \frac12\Dissipator{\Quadrature}
    +
    \frac12\Dissipator{\QuadratureP},
    \label{eq:heating_quadrature_decomposition}
\end{align}
and therefore corresponds to equal diffusion in both phase-space quadratures.
The last term describes bosonic number dephasing.
The dissipator reads
\begin{align}
    \Dissipator{\Operator}\DensityMatrix
    =
    \Operator\DensityMatrix\Operator^\dagger
    -
    \frac12
    \curlies{\Operator^\dagger\Operator,\DensityMatrix}.
\end{align}
The symmetry of the channels is crucial.
Loss changes parity,
\begin{align}
    \ParityOp\Annihilation\ParityOp=-\Annihilation,
\end{align}
and both heating jump operators are parity odd,
\begin{align}
    \ParityOp\Creation\ParityOp=-\Creation.
\end{align}
Number dephasing instead preserves parity,
\begin{align}
    [\NumberOperator,\ParityOp]=0.
\end{align}
Thus, loss and motional heating can populate the opposite parity sector, whereas number dephasing cannot.

\smsubsection{General expression}
For a general mixed state, the QFI is
\begin{align}
    \QFisher[
    \DensityMatrix,
    \Quadrature
    ]
    =
    2
    \sum_{m,n}
    A_{mn}
    \abs{
    G_{mn}
    }^2,
    \label{eq:qfi_noise_section}
\end{align}
where the state is decomposed as
\begin{align}
    \DensityMatrix
    =
    \sum_m
    \EigenVal_m
    \ket{\EigenVal_m}\bra{\EigenVal_m},
\end{align}
with shorthand notations
\begin{align}
    A_{mn}
    =
    \frac{
    \rounds{
    \EigenVal_m-\EigenVal_n
    }^2
    }{
    \EigenVal_m+\EigenVal_n
    },
    \qquad
    G_{mn}
    =
    \matrixelement{
    \EigenVal_m
    }{
    \Quadrature
    }{
    \EigenVal_n
    }.
\end{align}
Let $ \dot{\DensityMatrix} = \Liouville \DensityMatrix$ be the noise generator evaluated at $\DimTime=0$, and define
\begin{align}
    L_{mn}
    =
    \matrixelement{
    \EigenVal_m
    }{
    \Liouville
    \DensityMatrix
    }{
    \EigenVal_n
    }.
\end{align}
For a nondegenerate spectrum,
\begin{align}
    \dot{\EigenVal}_m
    =
    L_{mm},
\end{align}
while
\begin{align}
    \dot{G}_{mn}
    =
    \sum_{k\neq m}
    \frac{
    L_{mk}
    }{
    \EigenVal_m-\EigenVal_k
    }
    G_{kn}
    +
    \sum_{k\neq n}
    \frac{
    L_{kn}
    }{
    \EigenVal_n-\EigenVal_k
    }
    G_{mk}.
\end{align}
Differentiating Eq.~\eqref{eq:qfi_noise_section} gives
\begin{align}
    \DotQFisher
    &=
    2
    \sum_{m,n}
    \dot{A}_{mn}
    \abs{
    G_{mn}
    }^2+
    4
    \sum_{m,n}
    A_{mn}
    \mathrm{Re}
    \rounds{
    G_{mn}^{*}
    \dot{G}_{mn}
    },
    \label{eq:dotqfi_general}
\end{align}
where
\begin{align}
    \dot{A}_{mn}
    = &
    \frac{
    2
    \rounds{
    \EigenVal_m-\EigenVal_n
    }
    }
    {
    \rounds{
    \EigenVal_m+\EigenVal_n
    }
    }\rounds{
    \dot{\EigenVal}_m-\dot{\EigenVal}_n
    }
    - \frac{
    \rounds{
    \EigenVal_m-\EigenVal_n
    }^2
    }{
    \rounds{
    \EigenVal_m+\EigenVal_n
    }^2
    } \rounds{
    \dot{\EigenVal}_m+\dot{\EigenVal}_n
    }.
\end{align}
Consequently, $\DotQFisher$ is in general a complicated spectral functional of $\DensityMatrix$, $\Quadrature$, and the noise Liouvillian. 
In the following, we therefore identify special classes of states for which analytical expressions simplify.

\smsubsection{General bounds from motional heating and bosonic loss}
The strong fragility of parity-enhanced displacement sensing can be understood from general information-theoretic bounds.
Let us first consider phase-insensitive motional heating,
\begin{align}
    \partial_\DimTime\DensityMatrix
    =
    \DimHeatingRate
    \rounds{
    \Dissipator{\Annihilation}
    +
    \Dissipator{\Creation}
    }
    \DensityMatrix .
    \label{eq:diffusion_master_sm}
\end{align}
Using Eq.~\eqref{eq:heating_quadrature_decomposition}, this channel contains additive diffusion generated by the sensed quadrature $\Quadrature$ with rate $\DimHeatingRate/2$, together with an equal diffusion in the conjugate quadrature.
The directional component admits the Gaussian-randomization representation
\begin{align}
    \DensityMatrix(\DimTime)
    =
    \int\dd\xi\,
    \frac{
    e^{-\xi^2/(\DimHeatingRate\DimTime)}
    }{
    \sqrt{\pi\DimHeatingRate\DimTime}
    }
    e^{-\ImagUnit\xi\Quadrature}
    \DensityMatrix(0)
    e^{\ImagUnit\xi\Quadrature},
    \label{eq:diffusion_random_displacement}
\end{align}
followed by an additional parameter-independent diffusion in $\QuadratureP$.
The quantum Stam inequality~\cite{DePalma2017} and monotonicity of the QFI therefore give
\begin{align}
    \frac{1}{\QFisher(\DimTime)}
    \geq
    \frac{1}{\QFisher(0)}
    +
    \frac{\DimHeatingRate\DimTime}{2}.
    \label{eq:stam_main}
\end{align}
Equivalently,
\begin{align}
    \QFisher(\DimTime)
    \leq
    \frac{
    \QFisher(0)
    }{
    1+\DimHeatingRate\DimTime\QFisher(0)/2
    }.
    \label{eq:stam_alt}
\end{align}
Expanding Eq.~\eqref{eq:stam_alt} for short times gives
\begin{align}
    \DotQFisher
    \leq
    -
    \frac{\DimHeatingRate}{2}
    \QFisher^2.
    \label{eq:dotf_diffusion_bound}
\end{align}
Thus, phase-insensitive heating retains the universal quadratic fragility of displacement sensitivity.
The factor of one half relative to directional diffusion follows because only one half of the total heating channel acts through the sensed quadrature.
The characteristic decay time scales as
\begin{align}
    \DimTime_{\mathrm{frag}}
    \sim
    \frac{1}{\DimHeatingRate\QFisher}.
\end{align}
For Gaussian states, including squeezed thermal states, the bound is saturated.

A related bound can be formulated for bosonic loss,
\begin{align}
    \partial_\DimTime\DensityMatrix
    =
    \DimBosonLossRate
    \Dissipator{\Annihilation}\DensityMatrix .
\end{align}
The loss channel corresponds to attenuation through a beam-splitter interaction with vacuum and preserves a finite residual displacement susceptibility associated with the vacuum fluctuations themselves.
The displacement QFI satisfies
\begin{align}
    \frac{1}{\QFisher(\DimTime)}
    \geq
    \frac{e^{-\DimBosonLossRate\DimTime}}{\QFisher(0)}
    +
    \frac{1-e^{-\DimBosonLossRate\DimTime}}{4}.
    \label{eq:loss_bound}
\end{align}
Expanding Eq.~\eqref{eq:loss_bound} at short times gives
\begin{align}
    \DotQFisher
    \leq
    -
    \frac{\DimBosonLossRate}{4}
    \QFisher(\QFisher-4).
    \label{eq:loss_bound_short}
\end{align}
For highly sensitive states with $\QFisher\gg4$, Eq.~\eqref{eq:loss_bound_short} again reduces asymptotically to
\begin{align}
    \DotQFisher
    \lesssim
    -
    \frac{\DimBosonLossRate}{4}
    \QFisher^2.
\end{align}
These bounds show that strong displacement sensitivity and strong susceptibility to parity-breaking noise are not independent properties.
Instead, they are fundamentally linked through the phase-space structure responsible for the metrological enhancement itself.

\smsubsection{Analytical cases}

\smsubsubsection{Definite-parity states}
Given that the state has support only in one parity sector, we have
\begin{align}
    \QFisher
    =
    4\Trace{\DensityMatrix\Quadrature^2}.
    \label{eq:qfi_parity_identity_noise}
\end{align}
This identity applies to parity-filtered squeezed thermal states, parity-filtered phonon-added states, parity-filtered SG-shifted states, qcMAP-cats, and parity-selected KPO states, as long as the state remains confined to a single parity sector.
For pure number dephasing, parity is preserved, so Eq.~\eqref{eq:qfi_parity_identity_noise} remains applicable during infinitesimal dephasing evolution.
Using the adjoint Liouvillian,
\begin{align}
    \frac{\dd}{\dd\DimTime}\Tr(\DensityMatrix\Quadrature^2)
    =
    \DimBosonDephasingRate
    \Tr(\DensityMatrix\DissipatorAdjoint{\NumberOperator}\Quadrature^2),
\end{align}
and
\begin{align}
    \DissipatorAdjoint{\NumberOperator}\Operator
    =
    -\frac12[\NumberOperator,[\NumberOperator,\Operator]],
\end{align}
we obtain
\begin{align}
    \DissipatorAdjoint{\NumberOperator}\Quadrature^2
    =
    -2\rounds{\Annihilation^2+\Annihilation^{\dagger 2}}.
\end{align}
Hence
\begin{align}
    \DotQFisher_{\phi}
    =
    -8\DimBosonDephasingRate\mathrm{Re}\angles{\Annihilation^2}.
    \label{eq:dotqfi_parity_dephasing}
\end{align}
For number-diagonal parity states, $\angles{\Annihilation^2}=0$, and therefore
\begin{align}
    \DotQFisher_{\phi}=0.
\end{align}
For loss and motional heating, Eq.~\eqref{eq:qfi_parity_identity_noise} cannot be differentiated directly after the noise has acted because both channels populate the initially empty parity sector.
A strict diagnostic is the leakage rate into the opposite parity sector.
If the initial parity is $\pi$, then
\begin{align}
    \Gamma_{\mathrm{leak}}
    =
    \left.
    \frac{\dd}{\dd\DimTime}
    \Trace{\Projector_{-\pi}\DensityMatrix(\DimTime)}
    \right|_{\DimTime=0}.
\end{align}
For loss,
\begin{align}
    \Gamma_{\mathrm{leak}}^{\mathrm{loss}}
    =
    \DimBosonLossRate\angles{\NumberOperator},
    \label{eq:parity_leak_loss}
\end{align}
whereas for motional heating,
\begin{align}
    \Gamma_{\mathrm{leak}}^{\mathrm{h}}
    =
    \DimHeatingRate
    \rounds{
    \angles{\Creation\Annihilation}
    +
    \angles{\Annihilation\Creation}
    }
    =
    \DimHeatingRate\rounds{2\angles{\NumberOperator}+1}.
    \label{eq:parity_leak_diffusion}
\end{align}
For Fock-diagonal parity states, Eq.~\eqref{eq:qfi_parity_identity_noise} gives
\begin{align}
    \Gamma_{\mathrm{leak}}^{\mathrm{h}}
    =
    \frac{\DimHeatingRate}{4}\QFisher.
    \label{eq:diffusion_leak_equals_qfi}
\end{align}
The physical meaning becomes apparent from the spectral structure of the QFI.
For a perfect parity-definite state, the displacement operator connects the occupied support only to states with zero eigenvalue, so
\begin{align}
    \frac{(p_m-p_n)^2}{p_m+p_n}
    \xrightarrow{p_n\to0}
    p_m.
\end{align}
Parity-breaking noise immediately populates the opposite ladder.
Expanding for a small leaked population $q_n$ gives
\begin{align}
    \frac{(p_m-q_n)^2}{p_m+q_n}
    =
    p_m-3q_n+\mathcal O\rounds{q_n^2/p_m},
\end{align}
showing that the QFI decreases linearly with the leaked population.
For parity-filtered Fock-diagonal resources,
\begin{align}
    \QFisher=4\rounds{2\angles{\NumberOperator}+1},
\end{align}
while more generally $\QFisher\sim\angles{\NumberOperator}$ for the large resources considered here.
Bosonic loss then populates the opposite parity sector at a rate
\begin{align}
    \Gamma_{\mathrm{leak}}^{\mathrm{loss}}
    \sim
    \DimBosonLossRate\QFisher,
\end{align}
which gives the generic estimate
\begin{align}
    \DotQFisher_{\mathrm{loss}}
    \sim
    -\DimBosonLossRate\QFisher^2.
\end{align}
For motional heating, Eq.~\eqref{eq:diffusion_leak_equals_qfi} gives the same parametric relation,
\begin{align}
    \DotQFisher_{\mathrm{h}}
    \sim
    -\DimHeatingRate\QFisher^2,
\end{align}
consistent with the general bound in Eq.~\eqref{eq:dotf_diffusion_bound}.

\smsubsubsection{Fock-diagonal states}
For any Fock-diagonal state,
\begin{align}
    \DensityMatrix
    =
    \sum_n
    p_n
    \ket{n}\bra{n},
\end{align}
the QFI is exactly
\begin{align}
    \QFisher
    =
    4
    \sum_{n=0}^{\infty}
    \rounds{n+1}
    \frac{
    \rounds{
    p_n-p_{n+1}
    }^2
    }{
    p_n+p_{n+1}
    }.
    \label{eq:qfi_diagonal_noise}
\end{align}
This applies to thermal states, phonon-added thermal states, SG-shifted thermal states, and the corresponding parity-filtered variants before noise. 
For such states, number dephasing gives
\begin{align}
    \Dissipator{\NumberOperator}
    \DensityMatrix
    =
    0.
\end{align}
Therefore
\begin{align}
    \DotQFisher_{\phi}
    =
    0
\end{align}
for any initially Fock-diagonal state.

\smsubsubsection{Fock-diagonal states with definite parity}
A particularly important analytically tractable class is formed by states that are both diagonal in the Fock basis and confined to a single parity sector,
\begin{align}
    \DensityMatrix
    =
    \sum_{n\in s}p_n\ket n\bra n,
\end{align}
where $s$ denotes either the even or the odd ladder.
This class includes parity-filtered thermal states, parity-filtered phonon-added thermal states, and parity-filtered SG-shifted thermal states.
Since the displacement generator connects only opposite parities, the QFI is
\begin{align}
    \QFisher
    =
    4\Trace{\DensityMatrix\Quadrature^2}
    =
    4\rounds{2\angles{\NumberOperator}+1}.
    \label{eq:qfi_fock_diagonal_parity}
\end{align}
For zero-temperature loss, the state remains diagonal, but population is transferred to the opposite parity ladder.
Expanding the exact diagonal-state QFI to first order gives
\begin{align}
    \DotQFisher
    =
    -8\DimBosonLossRate
    \rounds{4\angles{\NumberOperator^2}-\angles{\NumberOperator}}.
    \label{eq:dotqfi_fock_parity_loss}
\end{align}
Hence, the loss susceptibility is controlled by the second number moment.
For number dephasing,
\begin{align}
    \Dissipator{\NumberOperator}\DensityMatrix=0,
\end{align}
and therefore
\begin{align}
    \DotQFisher=0.
    \label{eq:dotqfi_fock_parity_dephasing}
\end{align}
For phase-insensitive motional heating, both jump operators transfer population to the initially empty parity ladder.
The corresponding leakage rate is
\begin{align}
    \Gamma_{\mathrm{leak}}^{\mathrm{h}}
    =
    \DimHeatingRate\rounds{2\angles{\NumberOperator}+1}
    =
    \frac{\DimHeatingRate}{4}\QFisher.
    \label{eq:diffusion_leak_fock_parity}
\end{align}
The exact $\DotQFisher$ is generally not expressible only through a finite number of moments.
For a pure Fock state $\ket n$, however, direct expansion gives
\begin{align}
    \left.
    \frac{\partial\QFisher}{\partial\DimTime}
    \right|_{\DimTime=0}^{\mathrm{h}}
    =
    -\frac{\DimHeatingRate}{2}\QFisher_0^2
    =
    -8\DimHeatingRate\rounds{2n+1}^2.
    \label{eq:fock_diffusion_exact}
\end{align}
For mixed parity-filtered hot Fock states, we use Eq.~\eqref{eq:diffusion_leak_fock_parity} as the strict analytical leakage diagnostic and evaluate the full susceptibility numerically.

\smsubsubsection{Squeezed thermal states}
Squeezed thermal states form a Gaussian class for which loss and motional heating can be treated analytically.
Let
\begin{align}
    V_P
    =
    \rounds{2\ThermalOccupation+1}e^{-2\Squeeze}
\end{align}
be the variance of the quadrature conjugate to the displacement generated by $\Quadrature$.
The displacement QFI is
\begin{align}
    \QFisher_{\mathrm{sq}}
    =
    \frac{4}{V_P}
    =
    \frac{4e^{2\Squeeze}}{2\ThermalOccupation+1}.
\end{align}
Under zero-temperature loss,
\begin{align}
    V_P(\DimTime)
    =
    e^{-\DimBosonLossRate\DimTime}V_P(0)
    +
    \rounds{1-e^{-\DimBosonLossRate\DimTime}},
\end{align}
and therefore
\begin{align}
    \DotQFisher
    =
    4\DimBosonLossRate\frac{V_P-1}{V_P^2}.
    \label{eq:dotqfi_squeezed_loss}
\end{align}
For anti-squeezed displacement probes with $V_P<1$, this derivative is negative; for hot unsqueezed probes, loss can initially increase the QFI by cooling the state.
Under phase-insensitive motional heating,
\begin{align}
    V_P(\DimTime)
    =
    V_P(0)+2\DimHeatingRate\DimTime,
\end{align}
and hence
\begin{align}
    \DotQFisher
    =
    -\frac{8\DimHeatingRate}{V_P^2}
    =
    -\frac{\DimHeatingRate}{2}\QFisher^2.
    \label{eq:dotqfi_squeezed_diff}
\end{align}
Thus, squeezed thermal states saturate the heating bound in Eq.~\eqref{eq:dotf_diffusion_bound}.

\smsubsection{Effective two-branch decoherence model}
We now derive a minimal two-branch model for cat-like states whose dominant displacement sensitivity comes from coherence between two well-separated phase-space branches. 
The purpose of this model is not to replace the exact mixed-state QFI, but to obtain an analytical estimate for the initial noise susceptibility of the branch contribution.
Consider an effective two-dimensional branch subspace spanned by two approximately orthogonal states
\begin{align}
    \ket{+}
    \simeq
    \ket{\CohDisp},
    \qquad
    \ket{-}
    \simeq
    \ket{-\CohDisp},
\end{align}
with $\braket{+}{-}\simeq0$.
For real $\CohDisp$, the displacement generator has approximately diagonal matrix elements in this branch basis,
\begin{align}
    \matrixelement{+}{\Quadrature}{+}
    \simeq
    2\CohDisp,
    \qquad
    \matrixelement{-}{\Quadrature}{-}
    \simeq
    -2\CohDisp,
\end{align}
while the off-diagonal matrix element is exponentially small,
\begin{align}
    \matrixelement{+}{\Quadrature}{-}
    \simeq
    0.
\end{align}
Thus, within the branch subspace,
\begin{align}
    \Quadrature_{\mathrm{b}}
    \simeq
    2\CohDisp
    \PauliZ^{\mathrm{b}}.
    \label{eq:branch_generator_noise}
\end{align}
We model decoherence of the branch coherence by the effective density matrix
\begin{align}
    \DensityMatrix_{\mathrm{b}}
    =
    \frac12
    \begin{pmatrix}
    1 & \CatVisibility \\
    \CatVisibility & 1
    \end{pmatrix},
    \label{eq:branch_density_matrix}
\end{align}
where $0\leq \CatVisibility\leq1$ is the branch visibility. 
The eigenstates of Eq.~\eqref{eq:branch_density_matrix} are the even and odd branch superpositions,
\begin{align}
    \ket{\pm_{\mathrm{b}}}
    =
    \frac{
    \ket{+}\pm\ket{-}
    }{
    \sqrt2
    },
\end{align}
with eigenvalues
\begin{align}
    \lambda_{\pm}
    =
    \frac{
    1\pm\CatVisibility
    }{
    2
    }.
\end{align}
In this eigenbasis, the branch generator is off-diagonal,
\begin{align}
    \matrixelement{
    +_{\mathrm{b}}
    }{
    \Quadrature_{\mathrm{b}}
    }{
    -_{\mathrm{b}}
    }
    =
    2\CohDisp .
\end{align}
Substituting into the mixed-state QFI formula gives
\begin{align}
    \QFisher_{\mathrm{b}}
    &=
    2
    \sum_{\mu,\nu=\pm}
    \frac{
    \rounds{
    \lambda_\mu-\lambda_\nu
    }^2
    }{
    \lambda_\mu+\lambda_\nu
    }
    \left|
    \matrixelement{
    \mu_{\mathrm{b}}
    }{
    \Quadrature_{\mathrm{b}}
    }{
    \nu_{\mathrm{b}}
    }
    \right|^2
    \nonumber\\
    &=
    4
    \CatVisibility^2
    \left|
    \matrixelement{
    +_{\mathrm{b}}
    }{
    \Quadrature_{\mathrm{b}}
    }{
    -_{\mathrm{b}}
    }
    \right|^2
    \nonumber\\
    &=
    16
    \CohDisp^2
    \CatVisibility^2 .
    \label{eq:qfi_branch_visibility}
\end{align}
For a pure balanced cat, $\CatVisibility=1$, Eq.~\eqref{eq:qfi_branch_visibility} gives the familiar branch contribution $16\CohDisp^2$. 
For a fully incoherent mixture of the two branches, $\CatVisibility=0$, this branch contribution vanishes.

\smsubsubsection{Loss}
For zero-temperature loss, coherent amplitudes decay as
\begin{align}
    \CohDisp(\DimTime)
    =
    \CohDisp
    e^{
    -\DimBosonLossRate\DimTime/2
    }.
\end{align}
The off-diagonal coherent operator evolves as
\begin{align}
    &\ket{\CohDisp}\bra{-\CohDisp}
    \rightarrow
    \nonumber \\
    &\exp\squares{
    -
    2\CohDisp^2
    \rounds{
    1-e^{-\DimBosonLossRate\DimTime}
    }
    }
    \ket{
    \CohDisp e^{-\DimBosonLossRate\DimTime/2}
    }\bra{
    -\CohDisp e^{-\DimBosonLossRate\DimTime/2}
    }.
\end{align}
Then,
\begin{align}
    \CatVisibility_{\mathrm{loss}}(\DimTime)
    =
    \exp\squares{
    -
    2\CohDisp^2
    \rounds{
    1-e^{-\DimBosonLossRate\DimTime}
    }
    }.
\end{align}
Substitution into Eq.~\eqref{eq:qfi_branch_visibility} gives
\begin{align}
    \QFisher_{\mathrm{branch}}^{\mathrm{loss}}
    (\DimTime)
    \simeq
    16
    \CohDisp^2
    e^{-\DimBosonLossRate\DimTime}
    \exp\squares{
    -
    4\CohDisp^2
    \rounds{
    1-e^{-\DimBosonLossRate\DimTime}
    }
    }.
\end{align}
Therefore
\begin{align}
    \DotQFisher
    =
    -
    \DimBosonLossRate
    \rounds{
    1+4\CohDisp^2
    }
    \QFisher_{\mathrm{branch}}(0).
    \label{eq:dotqfi_cat_loss}
\end{align}

\smsubsubsection{Motional heating}
For phase-insensitive motional heating,
\begin{align}
    \frac{\dd\DensityMatrix}{\dd\DimTime}
    =
    \DimHeatingRate
    \rounds{
    \Dissipator{\Annihilation}
    +
    \Dissipator{\Creation}
    }
    \DensityMatrix .
\end{align}
Using Eq.~\eqref{eq:heating_quadrature_decomposition}, only the component acting through the branch-separation quadrature directly distinguishes the two branches.
For branches centered at $\pm\CohDisp$, the squared visibility therefore decays as
\begin{align}
    \CatVisibility_{\mathrm h}^2(\DimTime)
    \simeq
    \exp\rounds{-8\DimHeatingRate\CohDisp^2\DimTime}.
    \label{eq:heating_branch_visibility}
\end{align}
Since the branch separation is not reduced by this channel,
\begin{align}
    \DotQFisher
    =
    -8\DimHeatingRate\CohDisp^2\QFisher_{\mathrm{branch}}(0).
    \label{eq:dotqfi_cat_diff}
\end{align}
For a branch-dominated cat, $\QFisher_{\mathrm{branch}}(0)=16\CohDisp^2$, and Eq.~\eqref{eq:dotqfi_cat_diff} becomes
\begin{align}
    \DotQFisher
    =
    -\frac{\DimHeatingRate}{2}\QFisher_{\mathrm{branch}}^2,
\end{align}
which saturates the general heating bound.
During ECD generation, where $\CohDisp(\Time)=\Time$, integration over the growing branch separation gives
\begin{align}
    \CatVisibility_{\mathrm h}^2(\CatTime)
    =
    \exp\rounds{-\frac83\HeatingRate\CatTime^3},
    \label{eq:heating_visibility_ecd}
\end{align}
in agreement with the preparation-stage model used in Sec.~SV.

\smsubsubsection{Number dephasing}
Under number dephasing,
\begin{align}
    \DensityMatrix_{mn}(\DimTime)
    =
    \exp\rounds{
    -
    \frac{
    \DimBosonDephasingRate
    }{
    2
    }
    \rounds{
    m-n
    }^2
    \DimTime
    }
    \DensityMatrix_{mn}(0).
    \label{eq:number_dephasing_matrix_elements}
\end{align}
We now estimate the resulting decay of the branch coherence entering the effective cat model. 
Consider the off-diagonal branch operator
\begin{align}
    \DensityMatrix_{+-}
    =
    \ket{\CohDisp}\bra{-\CohDisp}.
\end{align}
Using the coherent-state expansion,
\begin{align}
    \ket{\pm\CohDisp}
    =
    e^{-\CohDisp^2/2}
    \sum_{n=0}^{\infty}
    \frac{
    (\pm\CohDisp)^n
    }{
    \sqrt{n!}
    }
    \ket{n},
\end{align}
the branch operator contains matrix elements
\begin{align}
    \matrixelement{m}{
    \DensityMatrix_{+-}
    }{
    n
    }
    \propto
    \frac{
    \CohDisp^{m+n}
    }{
    \sqrt{m!n!}
    }.
\end{align}
Equation~\eqref{eq:number_dephasing_matrix_elements} therefore suppresses the branch coherence according to the factor $
    \sim \exp[
    -
    \DimBosonDephasingRate
    (m-n)^2
    \DimTime/2
    ]$.
To characterize the resulting coherence decay, we consider the normalized Hilbert--Schmidt overlap
\begin{align}
    \CatVisibility_\phi(\DimTime)
    =
    \frac{
    \Trace{
    \DensityMatrix_{+-}^\dagger(0)
    \DensityMatrix_{+-}(\DimTime)
    }
    }{
    \Trace{
    \DensityMatrix_{+-}^\dagger(0)
    \DensityMatrix_{+-}(0)
    }
    }.
\end{align}
Substituting the coherent-state expansion gives
\begin{align}
    \CatVisibility_\phi(\DimTime)
    =
    \sum_{m,n}
    p_m
    p_n
    \exp\rounds{
    -
    \frac{
    \DimBosonDephasingRate
    }{
    2
    }
    (m-n)^2
    \DimTime
    },
\end{align}
where
\begin{align}
    p_n
    =
    e^{-\CohDisp^2}
    \frac{
    \CohDisp^{2n}
    }{
    n!
    }
\end{align}
is the Poisson distribution of the coherent branch.
For short times,
\begin{align}
    \CatVisibility_\phi(\DimTime)
    &=
    1
    -
    \frac{
    \DimBosonDephasingRate
    }{
    2
    }
    \angles{
    (m-n)^2
    }
    \DimTime
    +
    \mathcal O(\DimTime^2).
\end{align}
Since the two branches are statistically independent Poisson distributions with variance $\CohDisp^2$,
\begin{align}
    \angles{
    (m-n)^2
    }
    =
    2\CohDisp^2.
\end{align}
Therefore
\begin{align}
    \CatVisibility_\phi(\DimTime)
    \simeq
    \exp\rounds{
    -
    \DimBosonDephasingRate
    \CohDisp^2
    \DimTime
    }.
    \label{eq:cat_visibility_dephasing}
\end{align}
Using Eq.~\eqref{eq:qfi_branch_visibility}, we obtain
\begin{align}
    \DotQFisher
    \simeq
    -
    2
    \DimBosonDephasingRate
    \CohDisp^2
    \QFisher_{\mathrm{b}}(0).
\end{align}
This estimate captures the initial decay of branch coherence. 
It is not meant to replace the exact spectral QFI for mixed hot cats, but it provides the relevant scaling with cat size.
The effective branch model applies most directly to ECD hot cats, where the state is explicitly constructed as a coherent superposition of displaced thermal branches. 
It also applies to KPO-generated cats in the large-separation regime, provided the prepared state is well described by two semiclassical wells. 
For qcMAP and parity-selected KPO states, this branch-decoherence mechanism coexists with parity protection: loss and motional heating both suppress branch coherence and repopulate the initially empty parity sector.

\begin{figure}[ht!]
    \includegraphics[width=\linewidth]{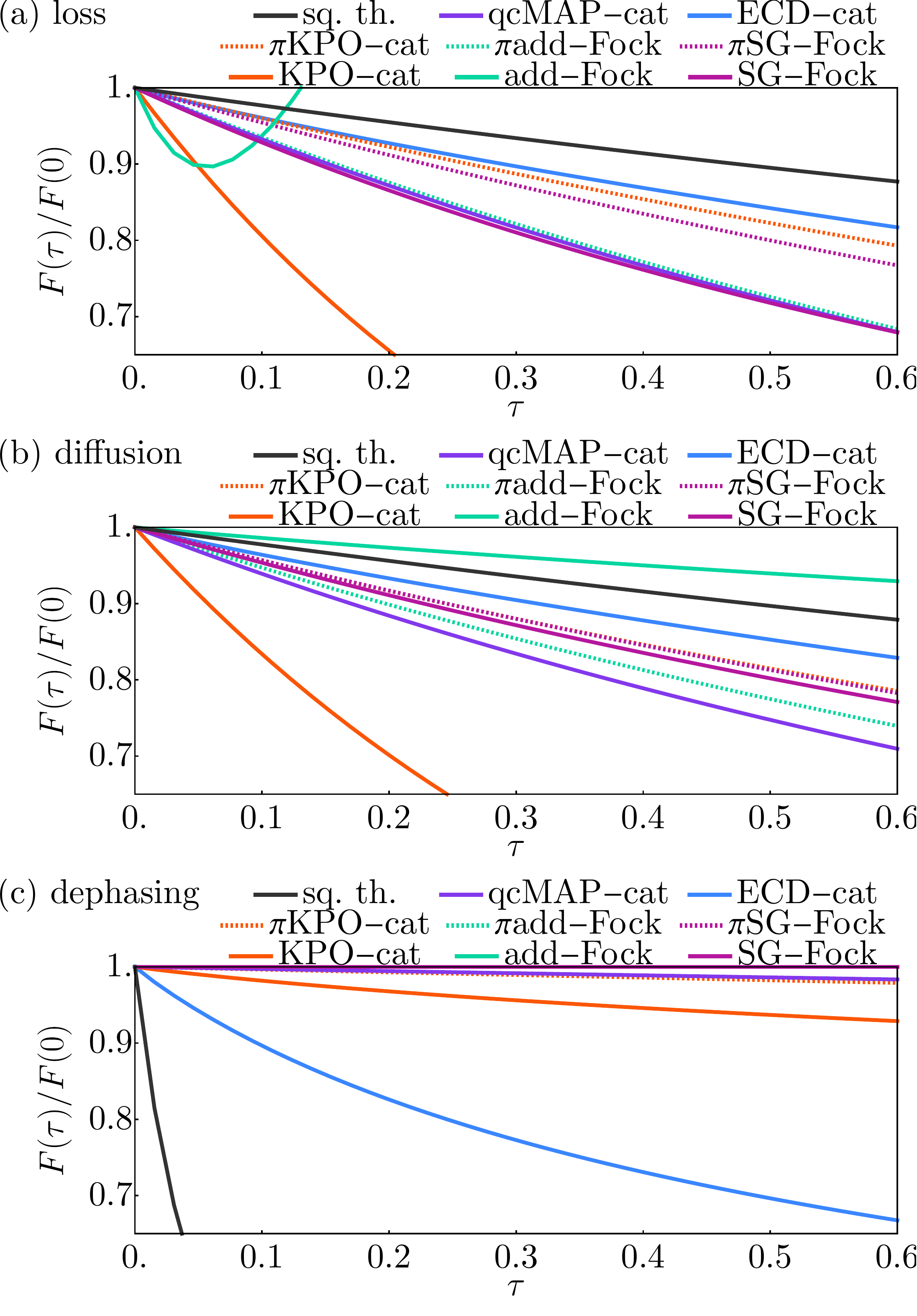}  
    \caption{Numerical verification of the noise susceptibility of different hot-state metrological resources. 
    Shown is the normalized displacement QFI, $\QFisher(\DimTime)/\QFisher(0)$, under (a) bosonic loss, (b) phase-insensitive motional heating, and (c) bosonic dephasing. All states were selected to have approximately equal initial QFI, $\QFisher(0)\approx25$, using the KPO-cat as the reference protocol.
    Squeezed thermal states were evolved within the Gaussian formalism, while all remaining states were simulated using full Fock-space Lindblad evolution.
    KPO-cats exhibit the fastest degradation under bosonic loss due to their comparatively large occupation at fixed initial QFI.
    Parity-filtered KPO-cats are substantially more robust than qcMAP-cats under both loss and heating, but remain less robust than ECD-cats.
    Parity filtering also strongly improves the robustness of add- and SG-Focks under all considered decoherence channels.
    Bosonic dephasing affects the compared resources qualitatively differently.
    Squeezed thermal states are highly sensitive because their metrological enhancement originates from phase-sensitive quadrature squeezing, whereas parity-filtered hot-Fock resources remain nearly insensitive due to their definite parity structure.
    Among the cat resources, ECD-cats show the strongest dephasing susceptibility.
  \label{FigSNoise}}
\end{figure}

\smsubsection{Numerical verification of noise susceptibility}
Here, we benchmark the analytical susceptibility predictions against full numerical Lindblad simulations for the different hot-state preparation protocols discussed in the main text. 
The goal of this analysis is not to optimize each protocol independently, but rather to compare the relative robustness of different metrological resources under identical decoherence channels and for comparable initial QFI.
The numerically compared states were selected such that their initial displacement QFI satisfied
\begin{align}
    \QFisher(0)
    \approx
    25,
\end{align}
up to small numerical mismatch originating from finite-Hilbert-space truncation and approximate parameter matching. 
The selected parameters and resulting exact initial properties are summarized in Table~\ref{tab:selected_states}. 
The ordinary KPO cat was used as the reference state, and parameters of the remaining protocols were chosen to approximately reproduce its initial QFI.
\begin{table}[t]
\centering
\begin{tabular}{lcc}
\hline
State & $\angles{\Creation\Annihilation}$ & $\QFisher(0)$ \\
\hline
ECD-cat & $2.38$ & $23.49$ \\
KPO-cat & $11.61$ & $22.97$ \\
$\pi$KPO-cat & $1.34$ & $23.04$ \\
qcMAP-cat & $1.60$ & $23.42$ \\
$\pi$add-Fock & $2.73$ & $25.87$ \\
$\pi$SG-Fock & $2.67$ & $25.33$ \\
add-Fock & $56.14$ & $22.66$ \\
SG-Fock & $9.00$ & $22.67$ \\
sq.~th. & $12.47$ & $22.97$ \\
\hline
\end{tabular}
\caption{
Initial properties of the numerically compared states used in Fig.~\ref{FigSNoise}.
}
\label{tab:selected_states}
\end{table}
The following families were compared: ECD-cats, qcMAP-cats, squeezed thermal states, add-Focks, $\pi$add-Focks, SG-Focks, $\pi$SG-Focks, KPO-cats, $\pi$KPO-cats.
For squeezed thermal states, the QFI and noisy evolution were evaluated entirely within the Gaussian formalism in order to avoid the extremely large Fock cutoffs required for converged thermal squeezing numerics at high occupations. 
All remaining states were evolved using full Fock-space Lindblad simulations.
The starting point is a KPO-cat that was generated using
\begin{align}
    \Kerr &= 1,
    &
    \frac{\CohDisp^2}{2\ThermalOccupation+1} &= 4,
    \nonumber \\
    \DimCatTime^\KPOLetter &= 100.
    &
    &
\end{align}
The corresponding Lindblad rates used for the three noise-channel checks are
\begin{align}
    \text{loss:}
    \qquad
    &
    \DimBosonLossRate / \Kerr=0.05,
    &
    \DimHeatingRate/ \Kerr&=0,
    &
    \DimBosonDephasingRate/ \Kerr&=0,
    \nonumber \\
    \text{heating:}
    \qquad
    &
    \DimBosonLossRate/ \Kerr=0,
    &
    \DimHeatingRate/ \Kerr&=0.01,
    &
    \DimBosonDephasingRate/ \Kerr&=0,
    \nonumber \\
    \text{dephasing:}
    \qquad
    &
    \DimBosonLossRate/ \Kerr=0,
    &
    \DimHeatingRate/ \Kerr&=0,
    &
    \DimBosonDephasingRate/ \Kerr&=0.05.
\end{align}
Figure~\ref{FigSNoise} shows the resulting normalized QFI dynamics,
\begin{align}
    \frac{
    \QFisher(\DimTime)
    }{
    \QFisher(0)
    }.
\end{align}
Several important qualitative trends emerge.

First, KPO-cats are highly fragile under bosonic loss. 
In Fig.~\ref{FigSNoise}(a), the KPO-cat curve shows the fastest relative decay among the compared states. 
This is consistent with the fact that the ordinary KPO reference state has a relatively large occupation,
$\angles{\Creation\Annihilation}\simeq 11.6$, so photon loss rapidly changes the state and degrades the QFI. 
Under motional heating, KPO-cats also decay rapidly, but the ordering is more nuanced because diffusion couples differently to coherent branch separation and local thermal structure. 
Parity-filtered KPO cats are much more robust in both loss and heating, reflecting the fact that they reach comparable initial QFI with much smaller occupation and a definite parity structure.

Second, parity-filtered KPO-cats show an intermediate robustness. 
Under both loss and motional heating, they are substantially more robust than qcMAP cats, but less robust than ECD cats. 
This indicates that restoring a definite parity sector improves the KPO construction, but does not make it as robust as the direct ECD branch-interference mechanism for the parameters considered here.
The comparison with qcMAP is particularly instructive. 
Both qcMAP and parity-filtered KPO states have definite parity, but qcMAP reaches the target QFI through a displaced parity-projected thermal state, whereas parity-filtered KPO reaches it through an excited-cat mixture with much smaller mean occupation. 
This leads to slower degradation for parity-filtered KPO under loss and heating. 
However, ECD cats retain the best robustness among the cat-like states in these two channels, despite not relying on parity projection.

Third, dephasing behaves qualitatively differently from loss and heating. 
Squeezed thermal states are extremely sensitive to dephasing because their metrological enhancement originates directly from phase-sensitive quadrature squeezing. 
In contrast, parity-filtered hot-Fock resources are nearly insensitive to bosonic dephasing, consistent with the analytical parity arguments derived previously.
Moreover, ECD-cats are much less robust than other cat resources under dephasing.

The numerics additionally confirm that parity filtering substantially improves the robustness of add- and SG-Focks under all considered decoherence channels. 
This agrees with the analytical expectation that parity selection removes adjacent-sector spectral overlap and restores the simplified parity-QFI structure.

Overall, the full numerical Lindblad simulations support the analytical susceptibility analysis developed in the previous subsections. 
In particular, they confirm that the relative robustness of hot-state metrological resources is governed not only by the total initial QFI, but also by how this QFI is distributed between coherent branch interference and local thermal or parity-protected contributions.

\smsubsection{General comparison and physical picture}

The analytical and numerical results together reveal that the noise susceptibility of hot-state metrological resources is governed not only by the total initial QFI, but also by the physical mechanism from which this QFI originates.
For cat-like resources, an important distinction is whether the metrological enhancement is dominated by coherent branch interference or by local parity-protected fluctuations. 
The effective two-branch model shows that coherent branch contributions scale as
\begin{align}
    \QFisher_{\mathrm{branch}}
    \sim
    16\CohDisp^2,
\end{align}
but are simultaneously highly sensitive to decoherence. 
Loss, motional heating, and number dephasing all suppress the branch visibility entering Eq.~\eqref{eq:qfi_branch_visibility}, leading to susceptibility scaling proportional to the cat size. 
In particular,
\begin{align}
    \DotQFisher_{\mathrm{loss}}
    &\sim
    -
    \DimBosonLossRate
    \CohDisp^2
    \QFisher,
    \\
    \DotQFisher_{\mathrm{diff}}
    &\sim
    -
    \DimHeatingRate
    \CohDisp^2
    \QFisher,
    \\
    \DotQFisher_{\phi}
    &\sim
    -
    \DimBosonDephasingRate
    \CohDisp^2
    \QFisher .
\end{align}
Thus, large coherent branch separation simultaneously enhances both the achievable sensitivity and the decoherence susceptibility.
This behavior is directly visible in the numerics for ECD- and qcMAP-cats. 
These protocols obtain large QFI predominantly from coherent branch interference and therefore exhibit strong sensitivity to bosonic loss and heating. 
Among the cat resources considered here, ECD-cats remain the most robust under these channels, while qcMAP-cats show substantially faster degradation. 
The difference reflects the fact that qcMAP states combine coherent cat structure with parity-projected thermal fluctuations, making them additionally sensitive to parity leakage generated by parity-breaking noise channels.
In contrast, parity-filtered hot-Fock resources derive their enhancement mainly from large quadrature fluctuations within a definite parity sector rather than from coherent branch interference. 
For such states, the corresponding dephasing susceptibility vanishes identically, $\DotQFisher_{\phi} = 0$, as long as the state remains Fock-diagonal.
This explains the dephasing robustness of parity-filtered add-Focks and SG-Focks.
At the same time, the same large quadrature variance responsible for the enhanced QFI also increases the sensitivity to motional heating,
\begin{align}
    \Gamma_{\mathrm{leak}}^{\mathrm{diff}}
    =
    \frac{
    \DimHeatingRate
    }{
    4
    }
    \QFisher ,
\end{align}
showing that diffusion acts directly on the amplified quadrature fluctuations themselves.
Thus, parity protection suppresses dephasing sensitivity, but does not protect against parity-breaking channels.
Parity-filtered KPO-cats occupy an intermediate regime between these two limiting behaviors. 
Parity projection restores a coherent cat contribution absent in ordinary hot KPO-cats, while simultaneously retaining a parity-filtered thermal contribution. 
Consequently, $\pi$KPO-cats are substantially more robust than qcMAP-cats under loss and heating, but less robust than ECD-cats.
The numerics show that they achieve comparable initial QFI with significantly smaller occupation than ordinary KPO-cats, which strongly improves their susceptibility to parity-breaking channels.
Ordinary hot KPO-cats exhibit qualitatively different behavior.
Their QFI is generated using a comparatively large occupation while possessing a weaker coherent branch structure. 
As a result, they are particularly fragile under bosonic loss, where excitation removal rapidly changes the state structure. 
This is clearly visible in Fig.~\ref{FigSNoise}(a), where ordinary KPO-cats show the fastest degradation among the compared resources.
Squeezed thermal states provide a complementary Gaussian benchmark.
Their QFI originates from phase-sensitive quadrature squeezing, and therefore their susceptibility is governed directly by the evolution of the squeezed quadrature variance.
Motional heating increases the squeezed variance linearly, leading to
\begin{align}
    \DotQFisher
    =
    -
    \frac{
    8
    \DimHeatingRate
    }{
    V_P^2
    },
\end{align}
while dephasing strongly suppresses the phase coherence required for squeezing itself.
This explains the particularly strong dephasing fragility observed numerically for squeezed thermal states.
Overall, the results demonstrate that there is no universal hierarchy of robustness among nonclassical metrological resources.
Different resources distribute their QFI differently between coherent branch interference, quadrature squeezing, thermal fluctuations, and parity-protected structure, and different noise channels couple selectively to these mechanisms.
The relevant operational question is therefore not only how large a QFI a state can achieve, but also which physical structure carries this QFI and how this structure transforms under experimentally relevant decoherence.

\smsection{Cooling versus direct ECD-cat preparation}
Section~SIV considered bosonic noise acting on an already prepared probe during sensing or storage.
Here we analyze decoherence during ECD preparation, when the relevant system is the joint spin--boson state.
ECD is additionally sensitive to ancilla spin dephasing, described by $(\DimSpinDephasingRate/4)\Dissipator{\PauliZ}$.
For circuit QED we consider photon loss, while for trapped ions we use the same phase-insensitive motional-heating channel $\DimHeatingRate[\Dissipator{\Annihilation}+\Dissipator{\Creation}]$ as in Sec.~SIV.
Bosonic number dephasing is not included in this minimal preparation model.
The cooling rate introduced below controls the preceding cooling stage and is not an additional decoherence channel during ECD generation.

We now formulate the operational question of whether cooling is advantageous before preparing an ECD-cat.
The point is not only to maximize the single-shot QFI of the final probe state, but to maximize the information accumulated within a fixed total laboratory time. 
A complete sensing cycle contains cooling, cat preparation, signal accumulation, and readout. 
We denote the corresponding durations by
\begin{align}
    \DimCycleTime
    =
    \DimCoolTime
    +
    \DimCatTime
    +
    \DimTimeSense
    +
    \DimTimeMeas .
\end{align}
If the total available experimental time is $\FullDimTime$, the number of repetitions is approximately
\begin{align}
    \NoRuns
    \simeq
    \frac{\FullDimTime}{\DimCycleTime}.
\end{align}
The quantum Cramér--Rao bound gives
\begin{align}
    \Delta \Estim
    \geq
    \frac{
    1
    }{
    \sqrt{
    \NoRuns
    \QFisher
    }
    },
\end{align}
so that the relevant operational figure of merit is the QFI accumulated per unit total protocol time,
\begin{align}
    \DimProtocolRate
    =
    \frac{
    \QFisher
    }{
    \DimCoolTime+\DimCatTime+\DimTimeSense+\DimTimeMeas
    }.
    \label{eq:full_protocol_rate}
\end{align}
We will use the simplified objective $\QFisher/(\DimCoolTime+\DimCatTime)$ that is justified only when $\DimTimeSense+\DimTimeMeas$ is either negligible compared with cooling and preparation, or fixed and sufficiently small that it does not qualitatively change the optimum. 
If $\DimTimeSense+\DimTimeMeas$ is fixed but not negligible, it should be retained as a constant offset in the denominator. 
This weakens the penalty for adding cooling time and can shift the optimum toward longer preparation, but it does not change the main conclusion whenever the dominant QFI contribution is insensitive to the cooled occupation.
We will use dimensionless times and rates normalized by the ECD coupling $\JCCoup$,
\begin{align}
    \CoolTime=\JCCoup\DimCoolTime,
    \qquad
    \CatTime=\JCCoup\DimCatTime,
    \qquad
    \TimeSense=\JCCoup\DimTimeSense,
    \qquad
    \TimeMeas=\JCCoup\DimTimeMeas,
\end{align}
and
\begin{align}
    \CoolingRate=\frac{\DimCoolingRate}{\JCCoup},
    \qquad
    \SpinDephasingRate=\frac{\DimSpinDephasingRate}{\JCCoup},
    \qquad
    \BosonLossRate=\frac{\DimBosonLossRate}{\JCCoup},
    \qquad
    \HeatingRate=\frac{\DimHeatingRate}{\JCCoup}.
\end{align}
Here $\CoolingRate$ describes relaxation during the cooling stage.
The remaining rates describe decoherence during the subsequent ECD preparation.
In particular, $\SpinDephasingRate$ refers to the ancillary spin and is distinct from the bosonic number-dephasing rate used in Sec.~SIV.
The dimensionless rate objective is then
\begin{align}
    \ProtocolRate
    =
    \frac{
    \QFisher
    }{
    \CoolTime+\CatTime+\TimeSense+\TimeMeas
    },
\end{align}
with the dimensional rate obtained by multiplying by $\JCCoup$.
During the cooling stage, the bosonic occupation relaxes according to
\begin{align}
    \frac{
    \dd n
    }{
    \dd\Time
    }
    =
    -
    \CoolingRate
    \rounds{
    n-\CoolingOccupation
    },
\end{align}
and therefore
\begin{align}
    \PostCoolingOccupation(\CoolTime)
    =
    \CoolingOccupation
    +
    \rounds{
    \ThermalOccupation-\CoolingOccupation
    }
    e^{-\CoolingRate\CoolTime}.
    \label{eq:post_cooling_occupation_rewrite}
\end{align}
For ideal ground-state cooling, one may set $\CoolingOccupation\simeq0$.
During ECD generation, the ideal rotating-frame Hamiltonian is
\begin{align}
    \frac{
    \Hamiltonian_{\ECDLetter}
    }{
    \hbar g
    }
    =
    \ImagUnit
    \PauliZ
    \rounds{
    \Creation-\Annihilation
    },
\end{align}
so that the branch displacement grows as
\begin{align}
    \CohDisp(\CatTime)=\CatTime .
\end{align}
In the large-separation regime, the ECD hot-cat QFI is well described by the branch-interference expression derived earlier,
\begin{align}
    \QFisher
    \simeq
    16
    \CatTime^2
    \CatVisibility^2
    +
    \frac{
    4
    }{
    2\PostCoolingOccupation(\CoolTime)+1
    }.
    \label{eq:qfi_ecd_cooling_general}
\end{align}
The first term is the coherent branch contribution, and the second is the local thermal contribution. 
This effective expression was checked against full numerical simulations for trapped-ion ECD parameters and captures the relevant optimization structure.
Spin dephasing during the ECD pulse suppresses the branch coherence. 
With the convention $\mathcal{L}_{\mathrm{s}}\DensityMatrix
    =
    \frac{
    \SpinDephasingRate
    }{
    4
    }
    \Dissipator{\PauliZ}
    \DensityMatrix$, the squared branch visibility acquires the factor
\begin{align}
    \CatVisibility_{\phi}^2
    =
    e^{-\SpinDephasingRate\CatTime}.
\end{align}
For circuit-QED implementations, the dominant bosonic channel during cat generation is typically photon loss, $\BosonLossRate\Dissipator{\Annihilation}$.
During the ECD pulse the two branches follow $\alpha_{\pm}(\Time)=\pm\Time$, so their separation is $\Delta\alpha(\Time)=2\Time$. 
The off-diagonal branch coherence under photon loss obeys
\begin{align}
    \frac{
    \dd\CatVisibility
    }{
    \dd\Time
    }
    =
    -
    \frac{
    \BosonLossRate
    }{
    2
    }
    \straights{
    \Delta\alpha(\Time)
    }^2
    \CatVisibility ,
\end{align}
hence
\begin{align}
    \CatVisibility_{\kappa}^2
    =
    \exp\rounds{
    -
    \frac{
    4
    }{
    3
    }
    \BosonLossRate
    \CatTime^3
    }.
\end{align}
The cQED preparation model is therefore
\begin{align}
    \QFisher_{\mathrm{cQED}}
    \simeq
    16
    \CatTime^2
    \exp\rounds{
    -
    \SpinDephasingRate\CatTime
    -
    \frac{
    4
    }{
    3
    }
    \BosonLossRate
    \CatTime^3
    }
    +
    \frac{
    4
    }{
    2\PostCoolingOccupation(\CoolTime)+1
    }.
    \label{eq:qfi_cqed_cooling_rewrite}
\end{align}
For trapped ions, the dominant bosonic decoherence during ECD generation is often motional heating. 
In the high-temperature diffusion limit, this is described by $\DimHeatingRate
    \rounds{
    \Dissipator{\Annihilation}
    +
    \Dissipator{\Creation}
    }$.
The corresponding branch-coherence decay is twice as strong as for zero-temperature loss in the above convention,
\begin{align}
    \CatVisibility_{\mathrm{h}}^2
    =
    \exp\rounds{
    -
    \frac{
    8
    }{
    3
    }
    \HeatingRate
    \CatTime^3
    }.
\end{align}
Hence, the trapped-ion preparation model becomes
\begin{align}
    \QFisher_{\mathrm{ion}}
    \simeq
    16
    \CatTime^2
    \exp\rounds{
    -
    \SpinDephasingRate\CatTime
    -
    \frac{
    8
    }{
    3
    }
    \HeatingRate
    \CatTime^3
    }
    +
    \frac{
    4
    }{
    2\PostCoolingOccupation(\CoolTime)+1
    }.
    \label{eq:qfi_ion_cooling_rewrite}
\end{align}
Equations~\eqref{eq:qfi_cqed_cooling_rewrite} and~\eqref{eq:qfi_ion_cooling_rewrite} show the essential structure of the optimization. 
Cooling changes only the local contribution through $\PostCoolingOccupation(\CoolTime)$.
During ECD preparation, spin dephasing suppresses the branch term in both platforms, while the accompanying bosonic channel is photon loss in circuit QED or motional heating in trapped ions. 
Therefore, whenever
\begin{align}
    16
    \CatTime^2
    \CatVisibility^2
    \gg
    \frac{
    4
    }{
    2\PostCoolingOccupation+1
    },
\end{align}
cooling gives only a small increase in QFI while increasing the denominator of Eq.~\eqref{eq:full_protocol_rate}. 
In this branch-dominated regime, direct hot-cat preparation is favored.
The simplified optimization is especially transparent when spin/control dephasing dominates,
\begin{align}
    \SpinDephasingRate
    \gg
    \BosonLossRate
    \quad\text{or}\quad
    \SpinDephasingRate
    \gg
    \HeatingRate .
\end{align}
Then,
\begin{align}
    \QFisher
    \simeq
    16
    \CatTime^2
    e^{-\SpinDephasingRate\CatTime}
    +
    \frac{
    4
    }{
    2\PostCoolingOccupation+1
    }.
\end{align}
Neglecting the local thermal term and keeping a fixed sensing/readout overhead $\Time_0
    =
    \TimeSense+\TimeMeas$,
the rate objective becomes
\begin{align}
    \ProtocolRate
    \propto
    \frac{
    \CatTime^2
    e^{-\SpinDephasingRate\CatTime}
    }{
    \CoolTime+\CatTime+\Time_0
    }.
    \label{eq:rate_with_fixed_overhead}
\end{align}
If $\Time_0\ll\CoolTime+\CatTime$, this reduces to the preparation-only objective. 
If $\Time_0$ is large and fixed, the penalty for increasing $\CoolTime$ or $\CatTime$ is weaker, but cooling is still useful only through the small local term. 
In the limiting case $\CoolTime=0$ and $\Time_0\ll\CatTime$, Eq.~\eqref{eq:rate_with_fixed_overhead} reduces to
\begin{align}
    \ProtocolRate
    \propto
    \CatTime
    e^{-\SpinDephasingRate\CatTime},
\end{align}
giving the optimal drive time,
\begin{align}
    \CatTime^{\star}
    \simeq
    \frac{
    1
    }{
    \SpinDephasingRate
    }.
\end{align}
In bosonic-decoherence-dominated regimes, the same reasoning gives algebraic optimal cat sizes. 
For cQED photon loss,
\begin{align}
    \ProtocolRate_{\mathrm{cQED}}
    \propto
    \CatTime
    \exp\rounds{
    -
    \frac{
    4
    }{
    3
    }
    \BosonLossRate
    \CatTime^3
    },
\end{align}
and therefore
\begin{align}
    \Time_{\mathrm{cQED}}^{\star}
    =
    \rounds{
    \frac{
    1
    }{
    4\BosonLossRate
    }
    }^{1/3}.
\end{align}
For trapped-ion motional heating,
\begin{align}
    \ProtocolRate_{\mathrm{ion}}
    \propto
    \CatTime
    \exp\rounds{
    -
    \frac{
    8
    }{
    3
    }
    \HeatingRate
    \CatTime^3
    },
\end{align}
which gives
\begin{align}
    \Time_{\mathrm{ion}}^{\star}
    =
    \rounds{
    \frac{
    1
    }{
    8\HeatingRate
    }
    }^{1/3}.
\end{align}
Typical circuit-QED hot-cat experiments operate with ECD or qcMAP protocols much faster than the cavity lifetime. 
For example, recently reported hot-cat measurements have used thermal occupations around $\langle \Creation \Annihilation  \rangle\simeq 2$, cat sizes around $\alpha=2.5$, and preparation/measurement times of order microseconds, shorter than the cavity relaxation time.
The heat bath can be disconnected before state preparation to avoid additional decoherence during the protocol. 
This places cQED ECD preparation naturally in a regime where spin/control coherence and finite pulse effects can be more restrictive than the initial thermal occupation. 
For trapped ions, a representative coupling scale is $\JCCoup/2\pi\sim 30\,\mathrm{kHz}$. 
Typical dimensionless parameters may lie in the ranges
\begin{align}
    \ThermalOccupation &\sim 1-20, 
    &
    \CoolingOccupation &\sim 0.01-0.1,
    \nonumber\\
    \CoolingRate &\sim 0.05-0.5,
    &
    \SpinDephasingRate &\sim 5\times10^{-4}-5\times10^{-2},
    \nonumber\\
    \HeatingRate &\sim 5\times10^{-6}-5\times10^{-3}.
\end{align}
In good traps, motional heating may be small, but it enters the branch coherence through the cubic factor $\CatTime^3$, so it becomes important for large cats.
For circuit QED, representative parameters are instead
\begin{align}
    g/2\pi &\sim 0.3\,\mathrm{MHz},
    &
    \ThermalOccupation &\sim 1-8,
    \nonumber \\
    \CoolingRate &\sim 0.05-5,
    &
    \SpinDephasingRate &\sim 0.03-0.1,
    \nonumber \\
    \BosonLossRate &\sim 5\times10^{-4}-5\times10^{-3}.
\end{align}
In this case, one often has $\SpinDephasingRate\gg\BosonLossRate$, so the spin-limited expression already captures the leading optimization.
To verify that the above conclusions are not artifacts of the effective branch-decoherence approximation, we additionally performed full numerical simulations of the complete Lindblad evolution during cooling and ECD preparation. 
The simulations include the cooling stage followed by ECD preparation with phase-insensitive motional heating and ancilla spin dephasing, and directly evaluate the final QFI of the generated sensing state.
Figure~\ref{Fig2} shows representative results for
\begin{align}
    \ThermalOccupation &= 10,
    &
    \CoolingOccupation &= 0,
    \nonumber \\
    \CoolingRate &= 1,
    &
    \SpinDephasingRate &= 10^{-2},
    \nonumber \\
    \HeatingRate &= 10^{-3}.
\end{align}
The preparation stage consists of a full numerical ECD evolution with variable cooling duration $\CoolTime$ and cat-generation duration $\CatTime$.
Panel~(a) shows the resulting QFI itself. 
As expected, the QFI grows monotonically with increasing cat-generation time because larger branch separation enhances the metrological sensitivity. 
The dependence on the cooling duration is comparatively weak, since cooling modifies mainly the local thermal contribution while leaving the dominant branch-interference term largely unchanged.
However, the relevant operational quantity is the QFI accumulated per unit total protocol time. 
Panel~(b) therefore shows $\QFisher/(\CoolTime+\CatTime)$.
A clear optimum emerges at finite preparation time, while the optimal cooling duration remains close to zero. 
Thus, although additional cooling slightly improves the single-shot QFI, the improvement is insufficient to compensate for the corresponding increase in protocol duration.
For the parameters used in Fig.~\ref{Fig2}, the optimum is quantitatively consistent with the heating-dominated estimate, which gives \(\CatTime^\star=5\) for \(\HeatingRate=10^{-3}\).
Importantly, the full numerical simulations reproduce the same qualitative behavior predicted by the analytical effective-decoherence model: in the branch-dominated regime, direct hot-cat preparation maximizes the metrological sensitivity achievable within a fixed total experimental runtime.
The resulting conclusion is not that cooling is never useful. 
Cooling becomes advantageous if the local thermal term is comparable to the branch term, if cooling changes the subsequent decoherence mechanism, if sensing or readout fidelity depends strongly on the thermal occupation, or if the initial occupation is so large that the large-separation ECD approximation breaks down. 
However, within the effective ECD-QFI model considered here, when branch coherence dominates the metrological gain and cooling affects only the initial thermal occupation, complete pre-cooling is generally not the optimal use of total experimental time.

\end{document}